\documentclass[nofootinbib,prd,showpacs]{revtex4-1}
\usepackage[utf8]{inputenc}
\usepackage{amsmath}
\usepackage{amsfonts}
\usepackage{amssymb}
\usepackage{graphicx}
\graphicspath{ {figures/} }
\usepackage[usenames,dvipsnames]{xcolor}
\usepackage{hyperref}   
\hypersetup{colorlinks=true, 
citecolor=MidnightBlue, linkcolor=MidnightBlue, urlcolor=Cyan}
\usepackage{tikz}
\usepackage{verbatim} 
\definecolor{purple}{rgb}{1,0,1}
\definecolor{lime}{HTML}{A6CE39} 

\newcommand{\orcidicon}{%
	\begin{tikzpicture}
	\draw[lime, fill=lime] (0,0) 
		circle [radius=0.16] 
		node[white] {{\fontfamily{qag}\selectfont \tiny ID}};
	\draw[white, fill=white] (-0.0625,0.095) 
		circle [radius=0.007];
	\end{tikzpicture}	\hspace{-2mm}
}

\newcommand\orcidManuel{{\href{https://orcid.org/0000-0001-8586-0285}{\orcidicon}}}

\begin{document}
\title{Coincident $f(\mathbb{Q})$ gravity: black holes, regular black holes and black bounces}


\author{Jos\'{e} Tarciso S. S. Junior.}
\email{tarcisojunior17@gmail.com}
\affiliation{Faculdade de F\'{i}sica, Programa de p\'{o}s-gradua\c{c}\~{a}o em F\'{i}sica,
Universidade Federal do Par\'{a}, CEP: 
66075-110, Belém, Par\'{a}, Brazil}

\author{Manuel E. Rodrigues\orcidManuel\!\!}
\email{esialg@gmail.com}
\affiliation{Faculdade de Ci\^{e}ncias Exatas e Tecnologia, 
Universidade Federal do Par\'{a}\\
Campus Universit\'{a}rio de Abaetetuba, 68440-000, Abaetetuba, Par\'{a}, 
Brazil}
\affiliation{Faculdade de F\'{\i}sica, Programa de P\'{o}s-Gradua\c{c}\~{a}o em 
F\'isica, Universidade Federal do 
 Par\'{a}, 66075-110, Bel\'{e}m, Par\'{a}, Brazil}

\date{19 May 2022; \LaTeX-ed \today}
\begin{abstract}

\bigskip

In this paper, we will use the coincident gauge to investigate new solutions of the $f(\mathbb{Q})$ theory applied in the context of black holes, regular black holes, and the black-bounce spacetime. 
For each of these approaches, we compute the linear solutions and the solutions with the constraint that the non-metricity scalar is zero. We also analyze the geodesics of each solution to interpret whether the spacetime is extensible or not, find the Kretschmann scalar to determine the regularity along spacetime, and in the context of regular black holes and black-bounce, we calculate the energy conditions. In the latter black-bounce case we realize that the null energy condition $(NEC)$, specifically the $NEC_1=WEC_1=SEC_1\leftrightarrow \rho+p_{r}\geq 0$, is satisfied outside the event horizon.

\end{abstract}
\maketitle
\def\HMS{{\scriptscriptstyle{\rm HMS}}}
\bigskip
\hrule
\tableofcontents
\bigskip
\hrule
\parindent0pt
\parskip7pt



\section{Introduction}
\label{sec1}

The theory of general relativity (GR), proposed by Albert Einstein in 1915, is the theory of gravity that currently shows the greatest agreement with the experimental tests by which it was challenged. In this context, we can cite some tests, for example, at the level of the solar system such as the precession of Mercury perihelion \cite{WILL2018} and the deflection of light by the curvature of spacetime near a massive body \cite{WILL2015}. In addition to these tests, other of his predictions have been confirmed recently, such as the detection of gravitational waves made in 2016 performed by the LIGO (Laser Interferometer Gravitational Wave Observatory) collaboration and Virgo~\cite{Abbott:2016blz,Abbott:2017vtc,LIGOScientific:2018mvr,TheLIGOScientific:2016pea,TheLIGOScientific:2016wfe,TheLIGOScientific:2016htt,Abbott:2020khf} and the very existence of black holes from their image construction, which was carried out by the international collaboration ``Event Horizon Telescope" (EHT) ~\cite{Akiyama:2019cqa, Akiyama:2019brx,Akiyama:2019sww,Akiyama:2019bqs,Akiyama:2019fyp,Akiyama:2019eap}.

In constructing GR, Einstein concluded that the concept of gravity is described by the curvature of space-time, with the mathematical structure described by Riemannian geometry~\cite{HILBERT, ALBERT}. On this occasion, the connection used to define the covariant derivative in GR is known as the Christoffel symbol or Levi-Civita connection. However, there are other objects described by non-Riemannian geometries that are capable of describing gravitational interaction beyond curvature~\cite{SHAPIRO},  which are: the torsion tensor $T^\alpha_{\phantom{\alpha}\mu\nu}$ and the non-metricity tensor $Q_{\alpha\mu\nu}$. Therefore, these tensors are the basis for establishing geometries used to mathematically structure other theories of gravity, but with a different description of the curvature of GR. These groups are grouped to define new tensors from their arrangements. The first is the contorsion tensor, this object is defined only in terms of the torsion tensor, while the second is the disformation tensor, defined with respect to the non-metricity tensor. Thus, these new quantities constitute, together with the connection of the GR, form the three components of the most general possible connection, the affine connection. Alternative theories of gravity can thus be developed from the composition of the affine connection.

Among the gravity models that can be described by affine geometry, there are first two specific cases of equivalence with GR. The first is known as the teleparallel equivalent of general relativity (TEGR), where gravitational effects are attributed only to the torsion tensor $T^\alpha_{\phantom{\alpha}\mu\nu}$~\cite{MOLLER, PELLEGRINI, CHO, HEHL}. Therefore, in the TEGR, the curvature $R^{\alpha}_{\phantom{\alpha}\mu\nu\beta}$ and the non-metricity tensor $Q_{\alpha\mu\nu}$ are assumed to be zero, and consequently this theory is characterized by a flat geometry. That is, in the teleparallel case, the parallel transported vectors do not change. Moreover, instead of presenting an action with the curvature scalar $R$ as in the theory GR, in the teleparallel theory we have an action written in terms of the torsion scalar plus a boundary term $(\mathbb{T}+B_T)$, which leads us to conclude that their functional actions are equivalent, this implies field equations written in terms of the torsion. 

While the second formulation, known as the teleparallel symmetric equivalent of general relativity (STEGR), contains the non-metricity tensor responsible for describing the gravitational interaction~\cite{NESTER}. So we have in this theory that the curvature $R^\alpha_{\phantom{\alpha}\mu\nu\beta}$ together with the torsion tensor $T_{\alpha}$ is zero. And as already mentioned in the teleparallel theory, STEGR is also developed by a non-Riemannian geometry of flat spacetime. It is attributed to this theory, as one of the simplest gravitational formulations, because the affine connection can be globally cancelled on a variety, thus allowing the covariant derivative to be described only in terms of the partial derivative, this choice of coordinate is called the coincident gauge~\cite{CGR}. 
These two approaches together with GR constitute the so-called ``The Geometrical Trinity of Gravity"~\cite{Trinity}.

However, despite the undoubted success of GR, this theory has some limitations in describing some phenomena of the Universe. At present, cosmological observations based on Type Ia Supernovae indicate that the Universe is in a phase of accelerated expansion~\cite{RIESS,PERLMUTTER}. But, this scenario lacks a convincing explanation from GR, i.e. in GR dynamics an exotic fluid exerting negative pressure, called dark energy (DE), is used, which has been inferred to ensure this scenario of the Universe. However there is no agreement on its origin. Observations indicate that DE governs approximately $70\%$ of the total energy density of the Universe~\cite{COPELAND,PEEBLES}.  It this context, that the scientific community uses new approaches to deepen the understanding in a more precise way, not only, the current expansion scenario of the Universe, but also dark matter, which has been postulated to explain the gravitational dynamics of galaxies~\cite{CAPOZZIELLO}.

On the other hand, in 1915 Karl Schwarzschild developed the first exact solution of the GR equations~\cite{Schwarzschild}, which later became known as black holes. This solution has some peculiarities, such as the structure known as the event horizon, a region from which, having reached it, not even a particle can escape~\cite{DINVERNO}. And it also predicts the existence of spacetime singularities at its center~\cite{STOICA}, a regime that makes GR inapplicable. However, Bardeen proposed a solution to the GR field equations in which the singularity does not exist~\cite{Bardeen}, this formulation is called regular black holes (RBH)~\cite{RODRIGUESSILVA}.  Using nonlinear electrodynamics coupled to GR, Beato and Garcia found exact solutions for the source of RBHs~\cite{BeatoGarcia1998,BRONNIKOV}. We also recommend~\cite{BeatoGarcia,BeatoGarcia1999b,BRONNIKOV2001,DYMNIKOVA}. Zhong-Ying Fan and Xiaobao Wang develop a method to obtain general solutions of black holes with electric or magnetic charges in General Relativity coupled with nonlinear electrodynamics~\cite{FAN,TOSHMATOV2018}. In their study, they concluded that the solutions of static, spherically symmetric black holes are regular throughout spacetime. Thermodynamics was also investigated, and consequently the first law of thermodynamics and Smarr's formula were found. And finally it was possible to generalize this procedure to obtain solutions for regular black holes with cosmological constant. Another class of black holes, namely those with rotation, is of great interest for astrophysical observations. Therefore, using the Newman-Janis algorithm method~\cite{NewmanJanis}, the authors develop generic solutions for rotating black holes in general relativity coupled to nonlinear electrodynamics\cite{TOSHMATOV,RODRIGUESJUNIOR}. In this work, the properties of rotating black hole solutions are discussed and some classes of geodesics are explored. It is also found that within the Cauchy horizon the weak and strong energy conditions are violated.

In addition, the theory GR predicts the existence of the spacetime structures capable of connecting two distant regions of the Universe or even of two different Universes, these objects are called wormholes~\cite{VISSER}. And they often require exotic matter that can violate energy conditions to remain stable~\cite{VISSER}.  The term wormholes, was established by Misner and Wheeler~\cite{WHEELER,MISNERWHEELER}. In 1935, Einstein and Rosen found the first wormhole solution using GR. Today this solution is known as the Einstein-Rosen bridge or Schwarzschild wormholes~\cite{EINSTEINROSEN}. Morris and Thorne were the ones who described a comprehensive study of traversable wormholes. Their work has contributed greatly to the spread of this subject~\cite{ MORRIS,MORRISTHORNE}.  A study of the quasi-normal modes, echoes, and shadow radii generated by wormholes from the Einstein-Maxwell-Dirac theories~\cite{SALCEDO,SALCEDO22} and in the Randall-Sundrum braneworld context~\cite{BRONNIKOV2003}, both without considering the use of exotic matter, was developed by~\cite{CHURILOVA2021}. And still in the context of wormholes in Einstein-Dirac-Maxwell theory without exotic matter in their structure, a study of epicyclic oscillatory motion was developed and from the frequency of orbital and epicyclic motion of test particles, the geodesics of quasiperiodic oscillations of high frequency models observed in active galactic nuclei were discussed~\cite{ZdenekVRBA}. In the work~\cite{KONOPLYA} solutions for asymmetric wormholes and soft matter were found, however, without satisfying the undesirable physical requirements of the Einstein-Dirac-Maxwell theory.

Recently Simpson and Visser proposed a formalism called black-bounce~\cite{visser2}. This structure of spacetime interpolates between Schwarzschild black holes and wormholes, and is described by a static spherically symmetric and asymptotically flat metric. Moreover, this new approach is free of singularities. In their construction, Simpson and Visser propose a metric consisting of a real parameter which, depending on its value, can describe the Schwarzschild metric, traversable wormholes, and the black-bounce. Thus, spacetime is regular if the bounce parameter is nonzero. After its development many works have been done on this subject, for example studies regarding regularity, causal structure and the energy conditions~\cite{LOBO}. Solutions with GR coupled with nonlinear electrodynamics and a scalar field~\cite{BronnikovWalia,Canate,ManuelMarcos}. Black-bounce solutions with charge and rotation~\cite{MAZZA,XUTANG,WU}. Weak and strong deflecting gravitational lenses in black-bounce spacetime~\cite{ZHANG}. A study of quasinormal modes was developed using the WKB method with Padé expansion~\cite{CHURILOVA}, in their development the authors conclude that the transition between black hole and wormhole is characterized by echoes. For some more discussion of this content see~\cite{YANG,FRANZIN,YANG2023}. And still in Simpson-Visser geometry a study of circular geodesic motion and epicyclic oscillatory motion was developed~\cite{VRBA}.

In view of the above, alternative approaches have been proposed to address the shortcomings of GR. There are, for example, the $f(R)$ theories~\cite{NOJIRI,SOTIRIOU,FELICE,Clifton}, which are generalizations of the GR theory, where instead of describing the action by a Ricci scalar $R$, arbitrary functions of this scalar are used. Starobinsky theory is a proposal that accounts for quantum corrections to an inflationary model by adding the $R^2$ term to the Einstein-Hilbert~\cite{Starobinsky} action. Other possibilities for theories of gravity are to use the trace of the energy-momentum tensor in the action, for example $f(R, \Theta)$~\cite{HARKO, JAMIL} and $f(R, \Theta, R_{ \mu\nu}\Theta^{\mu\nu})$~\cite{Sharif, ODINTSOV, AYUSO,SHARIF2018,ABCHOUYEH,GONÇALVES}, where $\Theta_{\mu\nu}$ represents the energy-momentum tensor, $\Theta$ is the trace of this tensor, and $R_{\mu\nu}$ is the Ricci tensor. Another approach to explain the expansion of the Universe, is to use functions of the Gauss-Bonnet invariant, $G$, in the gravitational action with $f(G)$~\cite{BAMBA,NOJIRI2010,RODRIGUES,HOUNDJO,AI}, $f(G,\Theta)$~\cite{ SHAMIR,AHMAD,YOUSAF} and also $f(R, G)$~\cite{NOJIRI2005,COGNOLA,DEFELICE,DEFELICE2009,ELIZALDE,ODINTSOV19}.

But just as there are generalizations $f(R)$, we also have their analog in teleparallel theories, i.e. instead of functions of the scalar $R$, arbitrary functions of the torsion scalar in the Lagrangian $f(\mathbb{T})$~\cite{CAI2016} are used. This construction represents a reasonable explanation for the late acceleration of the universe without including an exotic component~\cite{CAI2016}. This construction represents a reasonable explanation for the late acceleration of the universe without including an exotic component~\cite{CAI2016}. We also have other constructions from this extension, like the one with the trace of the energy-momentum tensor $f(\mathbb{T}, \Theta)$~\cite{KIANI,HARKO14,NASSUR,JUNIOR,GHOSH}, the inclusion of the Gauss-Bonnet analog invariant $T_G$, which we occasionally have the theory $f(\mathbb{T}, T_G)$~\cite{KOFINAS}. Another interesting solution of the extension $f(\mathbb{T})$, was applied in the context of black-bounce spacetime~\cite{Ednaldo}.

Recently, a new formulation of modified theories of gravity based on nonlinear functions of the non-metricity scalar $f(\mathbb{Q})$ has been proposed~\cite{CGR,JIMENEZ}. Given the limitations of GR and analogous to the modified theories $f({R})$ and $f(\mathbb{T})$, we have, for example, also in the theory $f(\mathbb{Q})$ a great interest in phenomenological applications from a new point of view, which is a promising basis for the study of its gravitational effects in general. On the level of cosmology of the Friedmann-Lemaitre-Robertson-Walker (FLRW) universe, the $f(\mathbb{Q})$ theory has the same interpretations as the $f(\mathbb{T})$ theory. However, these formulations show a different behavior at the level of cosmological perturbation theory~\cite{JIMENEZ}. Other cosmological approaches can be found in references~\cite{HARKO18,SOLANKI}.  An extension of this theory, the $f(\mathbb{Q},\Theta)$ theories, have also been developed~\cite{XU,Sahoo,GADBAIL}. Wormholes were applied on the basis of $f(\mathbb{Q})$~\cite{BANERJEE,MUSTAFA,HASSAN,PARSAEI}. The $f(\mathbb{Q})$ was used to study topological vacuum solutions described by a static and spherical symmetry~\cite{CALZA}. Relevant work has been done to study the evolution of primordial black holes in gravity $f(\mathbb{Q})$~\cite{CHANDA}. Another interesting study of black holes applied in this extension is the use of symmetries for spherically symmetric and stationary spacetimes~\cite{DAMBROSIO}. In the work~\cite{BAHAMONDE} the author explores the understanding of the coincident doctrine building on the reference~\cite{DAMBROSIO}. Spacetime configurations for the metric that are static and spherically symmetric are found and discussed.

So, our goal in this paper is to use the coincident gauge to obtain and investigate solutions of the gravity $f(\mathbb{Q})$ applied in the context of black holes, regular black holes and black-bounce space-time. The structure of this paper is organized as follows, in the~section~\eqref{sec2} we present the symmetric teleparallel theory, the $f(\mathbb{Q})$ theory and the coincident gauge. In section~\eqref{sec3} we discuss the metric we will use to obtain our solutions, as well as the components of some geometric quantities and hence the field equations of $f(\mathbb{Q})$. We show in section~\eqref{sec4} solutions of black holes in the $f(\mathbb{Q})$ theory, first in subsection~\eqref{subsec31} we find the linear solution. And subsequently in subsection~\eqref{subsec32} we discuss and analyze the case where the non-metricity scalar is zero. In section~\eqref{sec5} we develop the solutions for regular black holes, in the subsequent subsections we find the solutions for the symmetric teleparallel theory and for the null non-metricity scalar, respectively. In section~\eqref{sec6} we develop the solutions for black-bounce spacetime for the linear case, and also for the case where we consider $\mathbb{Q}=0$, and we also analyze the energy conditions. In one of the cases, due to the complexity of the solutions, we analyze the energy conditions graphically.   And finally, we present and discuss our conclusions and the prospects for future research that we intend to conduct in section~\eqref{sec7}.


\section{$f(\mathbb{Q})$ Gravity}
\label{sec2}
\par
The connection is a mathematical object that defines the covariant derivative and tells us how tensors should be transported through the manifold. In general relativity, established by Riemannian geometry, the connection is determined by a symmetric connection known as Christoffel symbol or Levi-Civita connection. However, there are two objects that make up a more general connection formed by its antisymmetric part and another component given by the relaxation of the metricity condition.
This connection is known as an affine connection, its explicit form is,
\begin{equation}
\Gamma_{\phantom{\beta}\mu\nu}^{\beta}=\left\{ _{\phantom{\beta}\mu\nu}^{\beta}\right\} +K_{\phantom{\beta}\mu\nu}^{\beta}+L_{\phantom{\beta}\mu\nu}^{\beta},
\label{conexion}
\end{equation}
where the first term indicates the Christoffel symbol, which is compatible with the metric,
\begin{equation}
\left\{ _{\phantom{\beta}\mu\nu}^{\beta}\right\} =\frac{1}{2}g^{\beta\alpha}\left(\partial_{\mu}g_{\nu\alpha}+\partial_{\nu}g_{\alpha\mu}-\partial_{\alpha}g_{\mu\nu}\right).\label{christoffel}
\end{equation}
The second term in turn is the antisymmetric component of the connection known as contorsion, $K_{\phantom{\beta}\mu\nu}^{\beta}$, expressed in terms of the torsion tensor $T_{\phantom{\beta}\mu\nu}^{\beta}=2\varGamma_{\phantom{\beta}\left[\mu\nu\right]}^{\beta}=-T_{\phantom{\beta}\nu\mu}^{\beta}$,
\begin{equation}
K_{\phantom{\alpha}\mu\nu}^{\beta}=\frac{1}{2}T_{\phantom{\alpha}\mu\nu}^{\beta}+T_{(\mu\phantom{\beta}\nu)}^{\phantom{(\mu}\beta},\label{tns_tor} 
\end{equation}
 and finally  we have the disformation  tensor $L_{\phantom{\beta}\mu\nu}^{\beta}$, 
\begin{equation}
L_{\phantom{\alpha}\mu\nu}^{\beta}=\frac{1}{2}Q_{\phantom{\alpha}\mu\nu}^{\beta}-Q_{(\mu\phantom{\beta}\nu)}^{\phantom{(\mu}\beta}=L_{\phantom{\beta}\nu\mu}^{\beta},\label{disf}
\end{equation}
which is defined in terms of the non-metricity tensor,
\begin{equation}
 Q_{\beta\mu\nu}\equiv \nabla_{\beta}g_{\mu\nu},   \label{tns_nmetric}
\end{equation}
where we now have that the covariant derivative  $\nabla_\mu$ is related to the affine connection \eqref{conexion}. 

To simplify the equations of motion, it is appropriate to define the superpotential, whose explicit form is,
\begin{equation}
P_{\phantom{\alpha}\mu\nu}^{\beta}=-\frac{1}{2}L_{\phantom{\alpha}\mu\nu}^{\beta}-\frac{1}{4}\left[\left(\tilde{Q}^{\beta}g_{\mu\nu}-Q^{\beta}\right)g_{\mu\nu}+\delta_{(\mu}^{\beta}Q_{\nu)}\right],\label{superpot}
\end{equation}
where
 $Q_{\alpha}=g^{\mu\nu}Q_{\alpha\mu\nu}=Q_{\alpha\phantom{\nu }\nu}^{\phantom{\alpha}\nu}$
e $\tilde{Q}_{\alpha}=g^{\mu\nu}Q_{\mu\alpha\nu}=Q_{\phantom{\nu}\alpha\nu}^{\nu}$ are the traces of the non-metricity tensor.
\par
So we can define the non-metricity scalar in a more compact form by contracting the non-metricity tensor \eqref{tns_nmetric} with the superpotential \eqref{superpot},
\begin{equation}
    \mathbb{Q}=-Q_{\beta\mu\nu}P_{\phantom{\alpha}}^{\beta\mu\nu}.\label{scalarQ}
\end{equation}
\par
The curvature tensor described by general relativity is defined by the Levi-Civita connection,
\begin{equation}
R_{\phantom{\beta}\mu\alpha\nu}^{\beta}=\partial_{\alpha}\Gamma_{\phantom{\beta}\nu\mu}^{\beta}-\partial_{\nu}\Gamma_{\phantom{\beta}\alpha\mu}^{\beta}+\Gamma_{\phantom{\beta}\alpha\rho}^{\beta}\Gamma_{\phantom{\rho}\nu\mu}^{\rho}-\Gamma_{\phantom{\beta}\nu\rho}^{\beta}\Gamma_{\phantom{\rho}\alpha\mu}^{\rho},    \label{tns_Riem}
\end{equation}
it is possible to perform a contraction in this tensor leading to the Ricci tensor,
\begin{equation}
    R_{\mu\nu}=R_{\phantom{\beta}\mu\beta\nu}^{\beta},
\end{equation}
the contraction of the Ricci tensor gives us the Ricci scalar,
\begin{equation}
  R= g^{\mu\nu}R_{\mu\nu}.
\end{equation}

Using the affine connection, the Riemann tensor \eqref{tns_Riem} can be rewritten from the following decomposition,   
\begin{equation}
R_{\phantom{\beta}\alpha\mu\nu}^{\beta}=\overset{C}{R}{}_{\phantom{\beta}\alpha\mu\nu}^{\beta}+\overset{C}\nabla_{\mu}V_{\phantom{\beta}\nu\alpha}^{\beta}-\overset{C}\nabla_{\nu}V_{\phantom{\beta}\mu\alpha}^{\beta}+V_{\phantom{\beta}\mu\rho}^{\beta}V_{\phantom{\rho}\nu\alpha}^{\rho}-V_{\phantom{\beta}\nu\rho}^{\beta}V_{\phantom{\rho}\mu\alpha}^{\rho},\label{Tns_Riem_Trans}  
\end{equation}
where $R_{\phantom{\beta}\alpha\mu\nu}^{\beta}$ is described in terms of the affine connection, $\overset{C}{R}{}_{\phantom{\beta}\alpha\mu\nu}^{\beta}$ and the derivative $\overset{C}\nabla$ are quantities that are related to the Christoffel symbol \eqref{christoffel}, and the tensor $V^\beta_{\phantom{\beta}\mu\nu}$ is given by, 
\begin{equation}
    V^\beta_{\phantom{\beta}\mu\nu}=K^\beta_{\phantom{\beta}\mu\nu}+L^\beta_{\phantom{\beta}\mu\nu}.
\end{equation}
\par
Furthermore, for a torsion-free connection $T_{\phantom{\alpha}\mu\nu}^{\beta}=0$, and from the appropriate contractions imposed on the Riemann tensor, the relation \eqref{Tns_Riem_Trans} reduces to,
\begin{equation}
R=\overset{C}{R}-\mathbb{{Q}}+\overset{C}\nabla_{\beta}\left(Q^{\beta}-\tilde{Q}^{\beta}\right),\label{scalar_Ric}
\end{equation}
where $\overset{C}{R}$ is the Ricci scalar expressed in terms of the Christoffel symbol.
\par
Starting from the teleparallel condition (TC) given by $R=0$, i.e. we will have a flat space-time that establishes the teleparallel geometries, we will find in this way a more general approach associating the Ricci scalar with the non-metricity scalar,
\begin{equation}
\overset{C}{R}=\mathbb{Q}-\overset{C}\nabla_{\beta}\left(Q_{\phantom{\beta}}^{\beta}-\tilde{Q}{}_{\phantom{\beta}}^{\beta}\right).
\label{scalar_Ric2}
\end{equation}
Consequently, this relation informs us that the non-metricity scalar differs from the Ricci scalar by a total derivative term or a boundary term,
\begin{equation}
B_Q=\overset{C}\nabla_{\beta}\left(Q^{\beta}-\tilde{Q}^{\beta}\right).\label{boundary}
\end{equation}
Then, the proposal of a gravitational theory known as the symmetric teleparallel equivalent of general relativity (STEGR), in which the gravitational interaction is described by means of the non-metricity tensor, is thus given by the following action,
\begin{equation}
S_{\text{STEGR}}=\int\sqrt{-g}d^{4}x\Big[\mathbb{Q}+2\kappa^2\mathcal{L}_m\Big]\label{action_Q}.
\end{equation}
we have that $\kappa^2=8\pi G/c^4$, where $G$ is the gravitational constant and $\mathcal{L}_{m}$ is the Lagrangian of the matter field. Note that from the relation \eqref{scalar_Ric2} it appears that the action of STEGR theory differs from the Einstein-Hiblert action of GR by a boundary term ($B_Q$), which means that STEGR is an equivalent formulation of general relativity.
\par
A non-linear extension of the STERG theory, is to use the following action,
\begin{equation}
S_{\rm f_Q}=\int\sqrt{-g}d^{4}x\Big[f\left(\mathbb{{Q}}\right)+2\kappa^2\mathcal{L}_{m}\Big]\label{action_f(Q)}.
\end{equation}
where $f(\mathbb{Q})$ can be an arbitrary function of the non-metricity scalar $\mathbb{Q}$. 
\par
The field equations of the theory $f(\mathbb{Q})$, are obtained by varying the action \eqref{action_f(Q)} with respect to the metric \cite{JIMENEZ},
\begin{equation}
\frac{2}{\sqrt{-g}}\nabla_{\alpha}\left(\sqrt{-g}f_{{Q}}\left(\mathbb{Q}\right)P_{\phantom{\alpha}\mu\nu}^{\alpha}\right)+\frac{1}{2}g_{\mu\nu}f(\mathbb{Q})+f_{{Q}}\left(\mathbb{Q}\right)\left(P_{{\mu{\alpha}{{\nu}}}}Q_{\nu}^{\phantom{\nu}\alpha\beta}-2Q_{\alpha\beta\mu}P_{\phantom{\alpha\beta}\nu}^{\alpha\beta}\right)=\kappa^2 \Theta_{\mu\nu},\label{eq_fie_f(Q)}
\end{equation}
where the momentum-energy tensor is given by $\Theta_{\mu\nu}$ and to simplify the equations we use the following notation $ f_{{Q}}\equiv \frac{\partial f(\mathbb{Q})}{\partial\mathbb{Q}}$.
\par
If we perform the functional variation of the equation \eqref{action_Q}, we find the linear equations of the STEGR theory, which are obtained when we do in \eqref{eq_fie_f(Q)} $f\mathbb{(Q)}=\mathbb{Q}$.
\par
We can rewrite the equations of motion \eqref{eq_fie_f(Q)} in a more convenient way according to the proposal of \cite{LIN, ZHAO, DAMBROSIO}, 
\begin{equation}
f_{{Q}}\left(\mathbb{Q}\right)G_{\mu\nu}-\frac{1}{2}g_{\mu\nu}\left(f(\mathbb{Q})-f_{Q}\left(\mathbb{Q}\right)\mathbb{Q}\right)+2f_{QQ}\left(\mathbb{Q}\right)P_{\phantom{\alpha}\mu\nu}^{\alpha}\partial_{\alpha}\mathbb{Q}=\kappa^2\Theta_{\mu\nu}.\label{eq_f(Q)}
\end{equation}
where $G_{\mu\nu}$ is the Einstein tensor described by \eqref{christoffel} and we denote $f_{QQ}\equiv  \frac{\partial}{\partial\mathbb{Q}}  \Big(\frac{\partial f(\mathbb{Q})}{\partial\mathbb{Q}} \Big)$.
\par
If we assume that the matter content behaves like a perfect fluid, where $\Theta^{\mu}_{\nu}=\text{diag}[\rho,-p_r,-p_t,-p_t]$ , we have that the energy conditions can be extended to the theory $f(\mathbb{Q})$.
\begin{eqnarray}
NEC_{1,2}=SEC_{1,2}=WEC_{1,2}=&&\;\;\;\rho+p_{r,t}\geq 0\,,\label{NEC}\\
SEC_3=&&\;\;\;\rho + p_{r} + 2p_t\geq 0\,,\label{SEC}\\
DEC_{1,2}=&&\;\;\;\rho \, - \mid p_{r,t}\mid \geq 0\, \  \  \text{ou} \ \ \rho\pm p_{r,t}\geq 0\,,\label{DEC}\\
DEC_{3}=WEC_3=&&\;\;\;\rho\geq 0\,.\label{WEC}
\end{eqnarray}
where we associate the radial and tangential components with indices 1 and 2, respectively.
\par
Starting from TC, which corresponds to a variety with a plane geometry characterizing a pure inertial connection, it is possible to perform an \emph{gauge} transformation of the linear group ${\rm GL }(4, {\cal \mathbb{R}})$ parameterized by $\Lambda_{\phantom{\alpha}\mu}^{\alpha}$ \cite{Trinity, JIMENEZ},
\begin{equation}
\Gamma_{\phantom{\alpha}\mu\nu}^{\alpha}=\left(\Lambda^{-1}\right)_{\phantom{\alpha}\beta}^{\alpha}\partial_{[\mu}\Lambda_{\phantom{\alpha}\nu]}^{\beta}.
\end{equation}
So we can write that the most general possible connection, through the general element of ${\rm GL}(4, {\cal \mathbb{R}})$, which is parametrized by the transformation of
$\Lambda_{\phantom{\alpha}\mu}^{\alpha}=\partial_{\mu}\xi^{\alpha}$, where $\xi^{\alpha}$ is an arbitrary vector field,
\begin{equation}
\Gamma_{\phantom{\alpha}\mu\nu}^{\alpha}=\frac{\partial x^{\alpha}}{\partial\xi^{\rho}}\partial_{\mu}\partial_{\nu}\xi^{\rho}. \label{coinc}
\end{equation}
This result shows us that the connection can be removed by a coordinate transformation.The transformation that results in the connection \eqref{coinc} being removed is called \emph{gage coincident} \cite{CGR}.
\par
Consequently, from the coincident gauge we have that the non-metricity tensor defined by \eqref{tns_nmetric} becomes, 
\begin{equation}
 Q_{\beta\mu\nu}\equiv \partial_{\beta}g_{\mu\nu}. \label{tns_nmetric2}
\end{equation}
In this manuscript we use the coincident gauge to compute our solutions. 

We will calculate the Kretschmann scalar $K = R^{\beta\mu\nu\alpha}R_{\beta\mu\nu\alpha}$ to ascertain the regularity of the space-time of our solutions. Then, from the expression~\eqref{Tns_Riem_Trans} this scalar takes the form of 

\begin{equation}
 K=\overset{C}{K}+\overset{L}{K}+2\overset{C}{R}{}^{\beta\alpha\mu\nu}\left(g_{\delta\beta}\nabla_{\mu}L_{\phantom{\beta}\nu\alpha}^{\delta}-g_{\delta\beta}\nabla_{\nu}L_{\phantom{\beta}\mu\alpha}^{\delta}+L_{\beta\mu\rho}L_{\phantom{\beta}\nu\alpha}^{\rho}-L_{\beta\nu\rho}L_{\phantom{\beta}\mu\alpha}^{\rho}\right).\label{K}
\end{equation}
where $\overset{C}{K}$ is the Kretschmann scalar described by GR and $\overset{L}{K}$ is the scalar depending uniquely on the distortion.

From this we can conclude that a spacetime is regular, i.e. without curvature singularities, if the Kretschmann scalar shows no divergences. 
So we calculate the Kretschmann scalar for each model. But since the curvature $R_{\phantom{\alpha}\beta\mu\nu}^{\alpha}=0$, which directly implies that the global Kretschmann scalar \eqref{K} is null, This gives us the following relation, $-\overset{L}{K}-2\overset{C}{R}{}^{\beta\alpha\mu\nu}\left(g_{\delta\beta}\nabla_{\mu}L_{\phantom{\beta}\nu\alpha}^{\delta}-g_{\delta\beta}\nabla_{\nu}L_{\phantom{\beta}\mu\alpha}^{\delta}+L_{\beta\mu\rho}L_{\phantom{\beta}\nu\alpha}^{\rho}-L_{\beta\nu\rho}L_{\phantom{\beta}\mu\alpha}^{\rho}\right)=\overset{C}{K}$. From this relation, we will then use $\overset{C}{K}$ to determine the regular space-time solutions that we will discuss in the next sections.

\subsection{Geodesics}

Let us start with the Lagrangian to find the equations of the geodesic
\begin{equation}
    {\cal L}=g_{\mu\nu}\dot{x}^{\mu}\dot{x}^{\nu}=\epsilon,
\end{equation}
where $\epsilon\rightarrow(-1,0,1)$ which indicate respectively the spacelike, lightlike and timelike geodesics. Then we will only have
\begin{equation}
    {\cal L}=\left[g_{00}\left(r\right)\dot{t}^{2}+g_{11}\left(r\right)\dot{r}^{2}+g_{22}\left(r\right)\dot{\theta}^{2}+g_{33}\left(r,\theta\right)\dot{\phi}^{2}\right]=\epsilon.
\end{equation}

The Euler-Lagrange equation
\begin{equation}
    \frac{d}{d \tau}\left(\frac{{\cal \partial L}}{\partial\dot{x}^{\mu}}\right)-\frac{{\cal \partial L}}{\partial x^{\mu}}=0,
\end{equation}
give us the following equations for the coordinates $t$, $ r$, $\theta$ and $\phi$
\begin{eqnarray}
    \begin{cases}
g_{00}(r)\dot{t}=E.\\
g_{11}\left(r\right)\ddot{r}-\left[g_{00}^{\prime}\left(r\right)\dot{t}^{2}+g_{11}^{\prime}\left(r\right)\dot{r}^{2}+g_{22}^{\prime}\left(r\right)\dot{\theta}^{2}+g_{33}^{\prime}\left(r,\theta\right)\dot{\phi}^{2}\right]=\delta\\
g_{22}\left(r\right)\ddot{\theta}+\dot{g}_{22}\left(r\right)\dot{\theta}-\mathring{g}_{33}\left(r,\theta\right)\dot{\phi}^{2}=0\\
g_{33}\left(r,\theta\right)\dot{\phi}=-l.
\end{cases}
\end{eqnarray}
where $E$ and $l$ are the energy and angular momentum, respectively. And in our notation we use that $(\, ^\prime \,)$ denotes the derivative with respect to the radial coordinate, $(\, \mathring{} \, )$ represents the derivative with respect to proper time.

Starting from the following expression
\begin{equation}
g_{\mu\nu}\dot{x}^{\mu}\dot{x}^{\nu}=g_{00}\left(r\right)\dot{t}^{2}+g_{11}\left(r\right)\dot{r}^{2}+g_{22}\left(r\right)\dot{\theta}^{2}+g_{33}\left(r,\theta\right)\dot{\phi}^{2}=\epsilon. \label{eq}
\end{equation}
Now lets substitute in the equation \eqref{eq} the following relations 
\begin{eqnarray}
  && g_{00}(r)\dot{t}=E,	\\
      &&  g_{33}\left(r,\theta\right)\dot{\phi}=-l,	
\end{eqnarray}
choosing the motion in the equatorial plane to be $\theta=\pi/2$, we get
\begin{equation}
    \dot{r}^{2}=g^{11}\left(r\right)\epsilon-\left(\frac{g^{11}\left(r\right)}{g_{00}(r)g_{33}(r,\theta)}\right)\left(g_{33}(r,\theta)E^{2}+g_{00}(r)l^{2}\right).
\end{equation}

Another very important expression that will help us understand the causal structure of our geometry in a meaningful way is the geodesics. Geodesics describe the paths on which the particles move. However, there are two types of geodesics, namely those that describe the shortest distance between two points on a curve, the so-called metric geodesics, and affine geodesics, which define the straightest possible curves in a geometry. In Riemannian geometry or general relativity, the distinction between these two geodesics is irrelevant because the two geodesics are equivalent. However, in more general geometries, as in the case of symmetric teleparallel theory, the two definitions are relevant because we have one equation for the geodesic described by the connection describing all of spacetime and another by the Levi-Civita connection. Then, in order to describe and better understand the causal structure from the solutions of each case we will treat, we will also calculate the equations of the geodesics, and we will use the geodesics described by the Levi-Civita connection because they are the most compatible with the observational data \cite{Will}. Since the equation of the geodesic with the global space-time connection is simply $\frac{d^2x^{\mu}}{d\tau^2}=0$, which clearly cannot describe the geodesics of black holes and black bounces, we must surely choose the equation of the geodesic with the Levi-Civita connection.


\section{Definition of the metric, geometric objects, and the equations of motion} \label{sec3}
\par
We will now point out the main components of the geometric quantities of the theory with which we are concerned in this manuscript. However, in order to simplify the discussion of the solutions we will deal with later in the context of black holes, regular black holes and black-bounce, we assume the most general possible static and spherically symmetric metric represented by,
\begin{equation}
ds^2=e^{a(r)}dt^2-e^{b(r)}dr^2-\Sigma(r)^2\left[d\theta^{2}+\sin^{2}\left(\theta\right)d\phi^{2}\right],\label{m_bb}
\end{equation}
where $a(r), b(r)$ and $\Sigma(r)$ are radial functions of the coordinates and are independent of time, and the determinant of the metric tensor is given by $g=-e^{a(r)+b(r)}\Sigma^4(r)\sin^2\theta$. Note that the metric constraint is determined by the value of $\Sigma(r)$. Therefore, for each case we will discuss later, we will give the corresponding metric from \eqref{m_bb}.
\par
With the metric defined, we will now compute the components of the tensor objects that establish the symmetric teleparallel theory. Thus, we have that the non-zero components of the non-metricity tensor \eqref{tns_nmetric} are, 
\begin{eqnarray}
&&Q_{100}	=e^{a\left(r\right)}a'\left(r\right),\quad Q_{111}	=-e^{b\left(r\right)}b'\left(r\right),\quad Q_{122}	=-2\Sigma(r)\Sigma'(r),\nonumber\\
&&Q_{133}	=-2\sin^2\left(\theta\right) \Sigma(r)\Sigma'(r),\quad Q_{233}	=-2\cos\left(\theta\right)\sin\left(\theta\right) \Sigma^{2}(r).\label{tnm}
\end{eqnarray}
where the symbol $(')$ represents the derivation according to the coordinate $r$. In possession of the nonmetricity tensor, we obtain the nonzero components of the disformation  tensor \eqref{disf},
\begin{eqnarray}
&& L_{\phantom{0}01}^0 =  L_{\phantom{0}10}^0  =-\frac{1}{2}a'\left(r\right),\quad L_{\phantom{1}00}^1  = -\frac{1}{2}e^{\left[a\left(r\right)-b\left(r\right)\right]}a'\left(r\right),\quad L_{\phantom{1}11}^1  =-\frac{1}{2}b'\left(r\right),\nonumber\\
&& L_{\phantom{1}22}^1  =  e^{-b\left(r\right)}\Sigma(r)\Sigma'(r),\quad L_{\phantom{1}33}^1  = e^{-b\left(r\right)}\sin\left(\theta\right)\Sigma(r)\Sigma'(r),\quad L_{\phantom{2}21}^2  =  L_{\phantom{2}12}^2=L_{\phantom{3}31}^3=L_{\phantom{3}13}^3 =-\frac{\Sigma'(r)}{\Sigma(r)},\nonumber\\
&& L_{\phantom{2}33}^2=\cos\left(\theta\right)\sin\left(\theta\right),\quad L_{\phantom{2}32}^3=L_{\phantom{2}23}^3=-\cot\left(\theta\right).\label{desf}
\end{eqnarray}
The nonzero components of the \eqref{superpot} are:
\begin{eqnarray}
&& P_{\phantom{0}01}^0	=P_{\phantom{0}10}^0=\frac{1}{8}\left[a'(r)-b'(r)-\frac{4\Sigma'(r)}{\Sigma(r)}\right];\,\, P_{\phantom{1}00}^1	=-\frac{e^{\left[a(r)-b(r)\right]}\Sigma'(r)}{\Sigma(r)};\,\, P_{\phantom{0}02}^0	=P_{\phantom{0}20}^0=P_{\phantom{1}12}^1	=P_{\phantom{1}21}^1=-\frac{1}{4}\cot\left(\theta\right);\nonumber\\
&&P_{\phantom{1}22}^1	=\frac{1}{4}e^{-b(r)}\Sigma(r)\left[\Sigma(r)a'(r)+2\Sigma'(r)\right];\, P_{\phantom{1}33}^1	=\frac{1}{4}e^{-b(r)}\sin^{2}(\theta)\Sigma(r)\left[\Sigma(r)a'(r)+2\Sigma'(r)\right];\, P_{\phantom{2}00}^2	=-\frac{e^{a(r)}\cot(\theta)}{2\Sigma^{2}(r)};\nonumber\\
&& P_{\phantom{2}11}^2 	=\frac{e^{b(r)}\cot(\theta)}{2\Sigma^{2}(r)};\quad P_{\phantom{2}12}^2	= P_{\phantom{2}21}^2 = P_{\phantom{3}13}^3 = P_{\phantom{3}31}^3 =-\frac{1}{8}\left[a'(r)+b'(r)\right];\quad P_{\phantom{3}32}^3 = P_{\phantom{3}23}^3=\frac{\cot(\theta)}{4}.\label{superp}
\end{eqnarray}
From the equation \eqref{scalarQ} the non-metricity scalar is given by,
\begin{equation}
\mathbb{Q}=-\frac{2e^{-b(r)}\Sigma'(r)\big[\Sigma(r)a'(r)+\Sigma'(r)\big]}{\Sigma^{2}(r)}.\label{scalarQ2}
\end{equation}
\par
The evolutionary equations, considering the metric \eqref{m_bb} and the equations \eqref{tnm}, \eqref{desf}, \eqref{superp} and \eqref{scalarQ2} are given below,
\begin{eqnarray}
\frac{1}{2}\Big[f(\mathbb{Q})-f_{Q}(r)\mathbb{Q}(r)\Big]-\frac{2e^{-b(r)}f_{QQ}(r)\mathbb{Q^{\prime}}(r)\Sigma^{\prime}(r)}{\Sigma(r)}+\phantom{=} &&\nonumber\\	\label{eq_field_BB_fQ}
+\frac{e^{-b(r)}f_{Q}(r)}{\Sigma ^{2}(r)} \Big[\Sigma(r)\left(b'(r)\Sigma'(r)-2\Sigma''(r)\right)+e^{b(r)}-\Sigma^{\prime 2}(r)\Big]	=\frac{1}{2}\kappa^{2}\rho,&&\\
 \frac{1}{2}f(\mathbb{Q})-\frac{1}{2}f_{Q}(r)\left[\mathbb{Q}(r)+\frac{2e^{-b(r)}\left(\Sigma(r)a'(r)\Sigma'(r)-e^{b(r)}+\Sigma^{\prime 2}(r)\right)}{\Sigma^{2}(r)}\right] 	=-\frac{1}{2}\kappa^{2}p_{r},&&\\
\frac{1}{2}e^{-b(r)}f_{QQ}(r)\cot(\theta)\mathbb{Q}^{\prime}(r)	=0,&&\\
\frac{f_{QQ}(r)\cot(\theta)\mathbb{Q}^{\prime}(r)}{\Sigma^{2}(r)}	=0,&&\\
 \frac{-e^{-b(r)}f(\mathbb{Q})}{4\Sigma(r)} \Big[2a'(r)\Sigma'(r)+\Sigma(r)\Big(2a''(r)-a'(r)b'(r)+a^{\prime 2}(r)\Big)-2b'(r)\Sigma'(r)+2e^{b(r)}\mathbb{Q}(r)\Sigma(r)+4\Sigma''(r)\Big]+&& \nonumber\\ 
+\frac{e^{-b(r)}}{4\Sigma(r)} \Big[-2f_{QQ}(r)\mathbb{Q^{\prime}}(r)\Big(\Sigma(r)a'(r)+2\Sigma'(r)\Big)+2e^{b(r)}f(\mathbb{Q})\Sigma(r)\Big]	=-\frac{1}{2}\kappa^{2}p_{t}.&&\label{eq_field_BB_fQ2}
\end{eqnarray}

The geodesic equation for the \eqref{m_bb} metric is given as
\begin{equation}
      \left(\frac{dr}{d s}\right)^2= - \frac{e^{-a(r)-b(r)} \left[l^2 e^{a(r)}+E^2 \Sigma^2 (r)\right]}{\Sigma ^2(r)} + e^{-b(r)}  \epsilon,\label{geodesic}
\end{equation}
where $l$ is angular momentum, $E$ is energy and $s$ is the affine parameter. 

In this manuscript we will analyze two cases. The first case deals with the linear theory, that is, with the equation \eqref{eq_f(Q)}, in general we see that we are clearly in the GR theory with $f(\mathbb{Q})=f_0 \mathbb{Q}+f_1$, with effective cosmological constant being given by $\Lambda_{eff}=\frac{1}{2}\frac{f_0}{f_1}$ and $\kappa_{eff}^2=\frac{\kappa^2}{f_1}$. The next case would be described with the non-metricity $\mathbb{Q}=0$, $f(\mathbb{Q})=f_0$ and $f_Q(\mathbb{Q})=f_1$, where we fall back to GR with the cosmological constant $\Lambda_{eff}=\frac{1}{2}{f_0}{f_1}$ and $\kappa_{eff}^2=\frac{\kappa^2}{f_1}$. The components of the non-metricity tensor \eqref{tnm}, show that for the case of the Minkowski metric where $a^\prime=b^\prime=0$, $\Sigma(r)^\prime=1$ and $\Sigma(r)=r$, reveal that not all components cancel each other out.  However, in the case where the geometry has only curvature, we have that all components of the Riemann tensor for the Minkowski metric are identically zero, and so are all other objects described from the curvature. We see that this is not analogously true for a space which has only the non-metric tensor. The same is true for the disformation  tensor \eqref{disf}, whose components are described by \eqref{desf}, and the same is true for the superpotential, whose components are given by \eqref{superp}. Thus, using the equation \eqref{scalarQ2} we see that the non-metricity scalar given by \eqref{scalarQ2bn} is not an invariant on diffeomorphism, this is in contrast to the invariant of GR, which is the curvature scalar, since we can already see for the Minkowski metric that the nonmetricity scalar diverges at the point where $r=0$. We note, however, that the definition of the equation \eqref{boundary} with zero curvature for $r=0$ has no divergence.

\section{Black hole solutions in the $f(\mathbb{Q})$ theory }
\label{sec4}

In this topic we will define the metric on which we will develop our black hole solutions. Then, using the equations of motion, we will discuss the black hole solutions for STEGR and for the case when the non-metricity scalar is zero.

The black hole solutions are obtained by applying $\Sigma(r)=r$ to \eqref{m_bb} so that the metric is described by ,
\begin{equation}
ds^2=e^{a(r)}dt^2-e^{b(r)}dr^2-r^2\left[d\theta^{2}+\sin^{2}\left(\theta\right)d\phi^{2}\right]\label{m_bn},
\end{equation}
where the determinant of the metric tensor is given by $g=-e^{a(r)+b(r)}r^4\sin^2\theta$.  
\par
From \eqref{scalarQ}, the non-metricity scalar for the metric in question \eqref{m_bn}, becomes
\begin{equation}
\mathbb{Q}=-\frac{2e^{-b(r)}\left[ra'(r)+1\right]}{r^2}.\label{scalarQ2bn}
\end{equation}
\par
Thus, assuming $\Sigma(r)=r$ in the \eqref{tnm}, \eqref{desf} and \eqref{superp} components, and using the scalar \eqref{scalarQ2bn} the evolutionary equations are written as,
\begin{eqnarray}
\frac{1}{2}[f(\mathbb{Q})-f_{Q}(r)\mathbb{Q}(r)]-\frac{2e^{-b(r)}f_{QQ}(r)\mathbb{Q^{\prime}}(r)}{r}+\frac{e^{-b(r)}f_{Q}(r)\left[rb'(r)+e^{b(r)}-1\right]}{r^2}	=&&\frac{1}{2}\kappa^{2}\rho,\label{eq_field_BN_fQ1}\\
\frac{1}{2}\left\{ f(\mathbb{Q})-f_{Q}(r)\left[\mathbb{Q}(r)-\frac{2e^{-b(r)}\left(ra'(r)-e^{b(r)}\right)}{r^{2}}\right]\right\} 	=&&-\frac{1}{2}\kappa^{2}p_{r},\\
\frac{1}{2}e^{-b(r)}f_{QQ}(r)\cot(\theta)\mathbb{Q}^{\prime}(r)	=&&0,\\
\frac{f_{QQ}(r)\cot(\theta)\mathbb{Q}^{\prime}(r)}{r^{2}}	=&&0,\\
-\frac{e^{-b(r)}f_{Q}(r)}{4r}\left[2a'(r)-2b'(r)+r\left(2a''(r)-a'(r)b'(r)+a'^2(r)+2e^{b(r)}\mathbb{Q}^{\prime}(r)\right)\right]+&&\nonumber	\\
+\frac{e^{-b(r)}}{4r}\left[2e^{b(r)}f(\mathbb{Q})r-2f_{QQ}(r)\mathbb{Q^{\prime}}(r)\left(ra'(r)+2\right)\right]	=&&-\frac{1}{2}\kappa^{2}p_{t}.\label{eq_field_BN_fQ}
\end{eqnarray}
\subsection{Black Holles in Symmetric teleparallel theory equivalent to general relativity}\label{subsec31}

With the intention of finding regular solutions, we consider the following functions $f(\mathbb{Q})=\mathbb{Q}+2\lambda$, $f_{Q}(r)=1$ and $f_{QQ}(r)=0$, where $\lambda$ is a constant that reduces the expressions (\ref{eq_field_BN_fQ1}-\ref{eq_field_BN_fQ}) to the following forms,
\begin{eqnarray}
\frac{e^{-b(r)}\left(rb'(r)+e^{b(r)}-1\right)}{r^{2}}+\lambda=&&\kappa^{2}\rho(r)\\
\frac{e^{-b(r)}\left(-ra'(r)+e^{b(r)}-1\right)}{r^{2}}+\lambda=&&-\kappa^{2}p_{r}(r)\\
\frac{e^{-b(r)}}{4r}\left\{ \frac{4\left(ra'(r)+1\right)}{r}+2b'(r)+2re^{b(r)}\left[2\lambda-\frac{2e^{-b(r)}\left(ra'(r)+1\right)}{r^{2}}\right]\right\} +\phantom{=}&&\nonumber\\
+\frac{e^{-b(r)}}{4r}\left\{ 2b'(r)-2ra''(r)-a'(r)\left(2-rb'(r)\right)-ra'^2(r)\right\} =&&-\kappa^{2}p_{t}(r)
\end{eqnarray}
\par
Because of the spherical symmetry of the Einstein solutions, we can make $p_r(r)=-\rho(r)$. So we assume that $a(r)=-b(r)$, so the evolutionary equations are as follows,
\begin{eqnarray}
\frac{e^{-b(r)}\left(rb'(r)+e^{b(r)}-1\right)}{r^{2}}+\lambda	=&&\kappa^{2}\rho(r)\\
\frac{e^{-b(r)}\left(rb'(r)+e^{b(r)}-1\right)}{r^{2}}+\lambda	=&&\kappa^{2}\rho(r)\\
\frac{e^{-b(r)}}{4r}\left[2rb''(r)+2re^{b(r)}\left(2\lambda-\frac{2e^{-b(r)}\left(1-rb'(r)\right)}{r^{2}}\right)-rb'^2(r)\right]+&&\\	
+\frac{e^{-b(r)}}{4r}\left[\left(2-rb'(r)\right)b'(r)+2b'(r)+\frac{4\left(1-rb'(r)\right)}{r}\right]	=&&-\kappa^{2}p_{t}(r)
\end{eqnarray}
\par
Now we have that the scalar of nonmetricity is, 
\begin{equation}
    \mathbb{Q}=\frac{2 e^{-b(r)} \left(r b'(r)-1\right)}{r^2}.\label{QBN}
\end{equation}
And the scalar of Ricci is,
\begin{equation}
    \overset{C}{R}=\frac{e^{-b(r)} \left[-r^2 b''(r)+r^2 b'(r)^2-4 r b'(r)-2 e^{b(r)}+2\right]}{r^2}.\label{RBN}
\end{equation}
\par
We know that symmetric teleparallel differs from general relativity except for one divergence term. So, to ensure the correctness of our solutions, we take the derivative of the traces of the non-metricity tensor for the black hole case. On this occasion we find the following expression for the boundary term \eqref{boundary},
\begin{equation}
    B_Q=\frac{e^{-b(r)} \left[r^2 b''(r)+r^2 \left(-b'(r)^2\right)+6 r b'(r)+2 e^{b(r)}-4\right]}{r^2}.\label{B_BN}
\end{equation}
it is therefore easy to verify that it follows from the equations \eqref{QBN}, \eqref{RBN} and \eqref{B_BN} in \eqref{scalar_Ric2} that our solutions are satisfied.
\par
For the black hole we consider that $b(r)=-\ln\left(-\frac{2 M}{r}-\frac{\lambda r^2}{3}+1\right)$, which by symmetry gives us the following components $\rho(r)=\frac{2 \lambda}{\kappa^{2}}$, and
$p_t(r)= -\frac{2 \lambda}{\kappa^{2}}$. Which recover the black hole solutions of general relativity. For these conditions the non-metricity scalar is described by 
\begin{equation}
    \mathbb{Q}=2 \lambda -\frac{2}{r^2}.\label{QBN1}
\end{equation}
And the Ricci scalar is,
\begin{equation}
    \overset{C}{R}=-4 \lambda.\label{RBN1}
\end{equation}
The total derivative term becomes,
\begin{equation}
    B_Q=6 \lambda -\frac{2}{r^2}.\label{B_BN1}
\end{equation}
Again using \eqref{scalar_Ric2} we verify that the expressions for the non-metricity scalars \eqref{QBN1} and Ricci, together with the term \eqref{B_BN1}, guarantee the veracity of our solutions.

As we commented above, the non-metricity scalar \eqref{QBN1} may not adequately indicate what happens to spacetime, but for this solution in particular, the Kretschmann $\overset{C}{K}$ is divergent at the point $r=0$, indicating a singularity at $r=0$, and this time the non-metricity scalar correctly indicates what happens to spacetime in the case of $r=0$ which diverges. Moreover, the analysis of geodesics for this approach leads to the same results as~\cite{STUCHLIKHLEDIK,FARAONI2015}.

\subsection{Black hole solution considering $\mathbb{Q}=0$}\label{subsec32}

For this case we will use $b(r)=-2\ln\left(1-\frac{2 M}{r}\right)$, where $\mathbb{Q}=0$ we have,

\begin{equation}
    -\frac{2 e^{-b(r)} \left(r a'(r)+1\right)}{r^2}=0,
\end{equation}
whose solution is,
\begin{equation}
a(r)=-\ln \left(\frac{r}{r_0} \right).
\end{equation}
Where $r_0$ is a constant.  
We will use in our solutions $r_0=1$. 

We note that the Kretschmann scalar for this case diverges in the limit from $r\to0$.
And the Kretschmann scalar analyzed at the horizon radius, is described by $1/4M^4$.  In the limit of $r\to\infty$ the Kretschmann scalar now shows an asymptotically flat behavior.

The equation of the geodesic \eqref{geodesic}, which is described by the parameters $a(r)$ and $b(r)$, is thus in this case dated by
\begin{equation}
   \dot{r}^2=- \frac{1 }{r} \left(1-\frac{2 M}{r}\right)^2 \left(E^2 r^2+\frac{l^2}{r}\right)+  \left(1-\frac{2 M}{r}\right)^2 \epsilon, \label{geod_bn}
\end{equation}
where $l$ is the angular momentum and $E$ is the energy.

We verify that the geodesic equation \eqref{geod_bn} for $r$ is very small. We note that the solution obtained for $r(s)$ diverges when we take the limit of $s\to-\infty$, which thus implies that this geodesic equation is extensible.
As well as for very large $r$, where we have that the geodesic equation is also extensible to future infinity. Now when we now consider this geodesic equation \eqref{geod_bn} for $r\to0$, we have as a solution a function of $r(s)$ described by the following approximation $r(s) \sim s^{1/3}$. This approximation shows us that the geodesic equation is not extensible to any real value. This result can be confirmed using the Kretschmann scalar $\overset{C}{K}$, which diverges in the limit $r(s)\rightarrow0$, which agrees with our previously discussed result. We also note that the solution of the geodesic equation, when evaluated at the horizon radius, presents $r(s)$ proportional to a constant, i.e., the geodesic equation is extensible at this horizon.

Let us assume the following functions $f(r)=0$, $f_{Q}(r)=1$ and $f_{QQ}(r)=0$. Now the components (\ref{eq_field_BN_fQ1}-\ref{eq_field_BN_fQ}) are given by, 
\begin{eqnarray}
 && \rho(r) = \frac{4 M^2}{\kappa ^2 r^4}, \\
&& \!\! p_r(r) = -\frac{1}{\kappa ^2r^2}, \\
&& p_t(r) = \frac{r^2-4 M^2}{4 \kappa ^2 r^4}.
\end{eqnarray}
Since we are doing $\mathbb{Q}=0$, we have that the Ricci scalar \eqref{scalar_Ric} becomes,
\begin{equation}
    \overset{C}{R}=-\frac{12 M^2+r^2}{2 r^4},
\end{equation}
where the divergence term \eqref{boundary} is given by,
\begin{equation}
    B_Q=\frac{12 M^2+r^2}{2 r^4}.
\end{equation}
Then, we have that $R$ and $B_Q$ satisfy \eqref{scalar_Ric2}.

Since we have treated a case where the non-metricity scalar $\mathbb{Q}=0$, we cannot treat this scalar to explore the divergences of the space, but to compensate, the boundary term $B_Q$ is divergent, indicating a singularity.
As we have already mentioned above, $\overset{C}{K}$ also diverges at this point.
\begin{equation}
    \overset{C}{K}=\frac{1296 M^4-1920 M^3 r+952 M^2 r^2-160 M r^3+17 r^4}{4 r^8}.
\end{equation}


\section{Regular Black Hole solutions in the $f(\mathbb{Q})$ theory } \label{sec5}

Regular black hole solutions are appreciated because they avoid divergences. In this context, we want to find regular STEGR solutions and Bardeen-type black hole solutions in this section.
Then we note that the equations found agree with the expressions of the general relativity theory. We also study solutions where we consider the non-metricity scalar to be zero, and consequently analyze the expressions of the energy conditions.

However, although we call them regular black holes, these solutions are singular, because of the functional form of the metric component.
\subsection{Regular Black Hole solutions at STTEGR}\label{subsec5A}
Regular black hole solutions in STTEGR can be obtained when we consider in (\ref{eq_field_BB_fQ}-\ref{eq_field_BB_fQ2}), $f(r)=\mathbb{Q}$, $f_{Q}(r)=1$ and $f_{QQ}(r)=0$, having the non-metricity scalar,
\begin{equation}
\mathbb{Q}= -\frac{2 e^{-b(r)} \left(r a'(r)+1\right)}{r^2}.\label{QBNR}
\end{equation}
\par
Therefore, the evolutionary equations in STTEGR are, 
\begin{eqnarray}
 \frac{e^{-b(r)} \left(r b'(r)+e^{b(r)}-1\right)}{r^2}=&&\kappa ^2 \rho (r), \\
 \frac{e^{-b(r)} \left(e^{b(r)}-r a'(r)-1\right)}{r^2}=&&-\kappa ^2 p_r (r), \\
 \frac{e^{-b(r)}}{4r}\left[2b'(r)-2a'(r)+r\left(a'(r)b'(r)-a'(r)^{2}-2a^{\prime\prime}(r)+2e^{b(r)}\mathbb{Q}(r)-2e^{b(r)}{\mathbb{Q^{\prime}}}(r)\right)\right]=&&-\kappa ^2p_t(r).
\end{eqnarray}
\par
In spherical symmetry, by making $p_r(r)=-\rho(r)$ and imposing $a(r)=-b(r)$, the evolutionary equations are reduced as, 
\begin{eqnarray}
 \frac{e^{-b(r)} \Big[r b'(r)+e^{b(r)}-1\Big]}{r^2}=&&\kappa ^2 \rho (r)\label{embnr} \\
 \frac{e^{-b(r)} \Big[r b'(r)+e^{b(r)}-1\Big]}{r^2}=&&\kappa ^2 \rho (r) \\
 \frac{e^{-b(r)} \Big[2 r b''(r)-r b'(r)^2+\left(2-r b'(r)\right) b'(r)+2 b'(r)\Big]}{4 r}=&&-\kappa ^2{p_t}(r)\label{embnr1}
\end{eqnarray}


\subsection{Solution for the regular Bardeen black hole}\label{subsec5B}

The Bardeen model is a particular case given by $b(r)=-\ln\left(1-\frac{2 M r^2}{\left(q^2+r^2\right)^{3/2}}\right)$, so the equations of motion from the expressions (\ref{embnr}-\ref{embnr1}) are, 
\begin{eqnarray}
 \frac{6 M q^2}{\kappa ^2 \left(q^2+r^2\right)^{5/2}}=&&\rho (r),\\
 \frac{M \left(9 q^2 r^2-6 q^4\right)}{\kappa ^2 \left(q^2+r^2\right)^{7/2}}=&&{p_t}(r).
\end{eqnarray}

Now, in this case, the non-merticity scalar takes the form of,
\begin{equation}
    \mathbb{Q}= \frac{12 M q^2}{\left(q^2+r^2\right)^{5/2}}-\frac{2}{r^2}.\label{scalarQBNR}
\end{equation}
\par
From the definition of the de Ricci scalar we obtain,
\begin{equation}
   \overset{C}{R}= \frac{6 M q^2 \left(r^2-4 q^2\right)}{\left(q^2+r^2\right)^{7/2}}.\label{scalarRBNR}
\end{equation}
The boundary term is described by,
\begin{equation}
B_Q=\frac{6 M q^2 r^2}{\left(q^2+r^2\right)^{7/2}}+\frac{36 M q^4}{\left(q^2+r^2\right)^{7/2}}-\frac{2}{r^2},\label{df1}
\end{equation}
therefore, we have that the equations \eqref{scalarQBNR}, \eqref{scalarRBNR} and \eqref{df1} satisfy the equation \eqref{scalar_Ric2}. We see that the non-metricity scalar \eqref{scalarQBNR} diverges in the limit $r\rightarrow 0$, but by subtracting \eqref{df1}, we see that the Ricci scalar \eqref{scalarRBNR} is always finite for any value of the coodenate $r$.

This solution presents a divergence due to the metric function $a(r)$, which was obtained from the metric function $b(r)$ from the Bardeen model and through $\mathbb{Q}=0$. This is reflected in other quantities like the geodesics and the curvature scalar.

Calculating the value of the Kretschmann $\overset{C}{K}$ for the Bardeen metric, we see that the Kretschmann scalar is finite for any value of the radial coordinate, which proves that this spacetime is regular.
Also here we find out that only with the analysis of the non-metricity scalar it is not possible to determine what happens with the space-time.


\subsection{First solution of regular black holes with the zero non-metricity scalar, $\mathbb{Q}=0$} \label{subsec5C}

Suppose that $b(r)=-2\ln\big[1 - 2Mr^2/(r^2 + q^2)^{3/2}\big]$ and in this way we can find the solution for $a(r)$ from the condition $\mathbb{Q}=0$. In the case of regular black holes, the non-metricity scalar \eqref{scalarQ}, becomes
\begin{equation}
0=-\frac{2 e^{-b(r)} \big[r a'(r)+1\big]}{r^2},
\end{equation}
then we find the following solution, 
\begin{equation}
a(r)=-\ln \left(\frac{r}{r_0}\right)
\end{equation}
where $r_0$ is a constant. We will use $r_0=1$. 

We also compute the Kretschamnn scalar in terms of the metric functions $a(r)$ and $b(r)$ of this case, so we see that this scalar diverges when we evaluate the limit $r\to0$. Evaluating the Kretschmann scalar on the horizon radius, we see that this scalar depends only on the constants which are present in this case. And when we analyze the Kretschmann scalar for $r\to \infty$, we notice an asymptotically flat behavior. 

The equation of the geodesic for this case becomes,
\begin{equation}
    \dot{r}^2=-\frac{1}{r}\left(E^2 r^2+\frac{l^2}{r}\right) \left(1-\frac{2 M r^2}{\left(q^2+r^2\right)^{3/2}}\right)^2+\epsilon  \left(1-\frac{2 M r^2}{\left(q^2+r^2\right)^{3/2}}\right)^2, \label{geod_bnr}
\end{equation}
where $l$ is the angular momentum and $E$ is the energy.

Checking the geodesic equation \eqref{geod_bnr} for $r\to\infty$, we note that the solution found for $r(s)$ is extensible when we make $s\to \infty$.
From this we conclude that the geodesic equation is extensible to this limit. 
And when we consider $r $ to be very small, we also find that the geodesic equation is extensible.
When we evaluate the geodesic equation at the horizon radius, we obtain a solution for $r(s)\sim s^2$, indicating that this geodesic equation is extensible.

To simplify the equations we will adopt that $f(\mathbb{Q})=0$, $f_{Q}(r)=1$ and $f_{QQ}(r)=0$. From the evolutionary equations, but taking into account the symmetry in question, it follows that the components of the momentum-energy tensor are written in the form,
\begin{eqnarray}
&& \kappa ^2  \rho (r)=\frac{4 M \left(M \left(r^4-5 q^2 r^2\right)+3 q^2 \left(q^2+r^2\right)^{3/2}\right)}{ \left(q^2+r^2\right)^4},\label{embnr21} \\
&& \kappa ^2p_r(r)=-\frac{1}{r^2}, \\
&& \kappa ^2p_t(r)=  -\frac{ M^2 r^2 \left(r^2-5 q^2\right)}{\left(q^2+r^2\right)^4}-\frac{ 3M q^2}{\left(q^2+r^2\right)^{5/2}}+\frac{1}{4r^2} .\label{embnr22}
\end{eqnarray}
Performing the limit $r\to 0$ we have that $\rho(r)\to 6 M/q^3\kappa^2$, $p_r\to-\infty$ and $p_t\to+\infty$. For the $r\to\infty$ limit, on the other hand, all components are zero. 
\par
Since we are considering $\mathbb{Q}=0$, we have that the Ricci scalar is now given by, 
\begin{equation}
    \overset{C}{R}=\frac{6 M \left(5 M q^2 r^2 \sqrt{q^2+r^2}-M r^4 \sqrt{q^2+r^2}-3 q^2 \left(q^2+r^2\right)^2\right)}{\left(q^2+r^2\right)^{9/2}}-\frac{1}{2 r^2}.\label{RBNR}
\end{equation}
And the term of divergence \eqref{boundary} reads as, 
\begin{equation}
    B_Q=\frac{6 M \left(-5 M q^2 r^2 \sqrt{q^2+r^2}+M r^4 \sqrt{q^2+r^2}+3 q^2 \left(q^2+r^2\right)^2\right)}{\left(q^2+r^2\right)^{9/2}}+\frac{1}{2 r^2}.\label{B_BNR}
\end{equation}
And again, it is clear to see, in conjunction with the adopted non-metricity scalar, that the expressions \eqref{RBNR} and \eqref{B_BNR} satisfy \eqref{scalar_Ric2}. 
\par
Starting from the equations of motion (\ref{embnr21}-\ref{embnr22}) and the expressions of the energy conditions (\ref{NEC}-\ref{WEC}) for outside the event horizon, i.e. in regions where $t$ is the time component, we have that the energy conditions are now given by,
\begin{eqnarray}
&&NEC_{1}=SEC_{1}=WEC_{1}=	\frac{1}{\kappa^{2}}\left[\frac{4 M \left(M \left(r^4-5 q^2 r^2\right)+3 q^2 \left(q^2+r^2\right)^{3/2}\right)}{\left(q^2+r^2\right)^4}-\frac{1}{r^2}\right]\ge0\\
&&NEC_{2}=SEC_{2}=WEC_{2}=	\frac{1}{4\kappa^{2}}\left[\frac{12 M^2 r^2 \left(r^2-5 q^2\right)}{\left(q^2+r^2\right)^4}+\frac{36 M q^2}{\left(q^2+r^2\right)^{5/2}}+\frac{1}{r^2}\right]\ge0,\\
&&SEC_{3}=		\frac{1}{2\kappa^{2}}\left[\frac{4 M^2 r^2 \left(r^2-5 q^2\right)}{\left(q^2+r^2\right)^4}+\frac{12 M q^2}{\left(q^2+r^2\right)^{5/2}}-\frac{1}{r^2}\right]\ge0,\\
&&DEC1=		\frac{1}{\kappa^{2}}\left[ \frac{4 M \left(M \left(r^4-5 q^2 r^2\right)+3 q^2 \left(q^2+r^2\right)^{3/2}\right)}{\left(q^2+r^2\right)^4}+\frac{1}{r^2}\right]\ge0,\\
&&DEC_{2}=	\frac{1}{4\kappa^{2}}\left[ \frac{20 M^2 r^2 \left(r^2-5 q^2\right)}{\left(q^2+r^2\right)^4}+\frac{60 M q^2}{\left(q^2+r^2\right)^{5/2}}-\frac{1}{r^2}\right]\ge0,\\
&&DEC_{3}=WEC_{3}=\frac{4 M \left(M \left(r^4-5 q^2 r^2\right)+3 q^2 \left(q^2+r^2\right)^{3/2}\right)}{\kappa ^2 \left(q^2+r^2\right)^4}\geq0.
\end{eqnarray}

Analyzing our energy conditions above, we find that $NEC_1$, $SEC_3$ and $DEC$ are violated when we evaluate these solutions for very small $r$. And only the $NEC_2$ energy condition is satisfied. The energy conditions $NEC_1$, $SEC_3$, $DEC_2$ and $DEC_3$, are violated when we evaluate these energy conditions for very large $r$. While the energy conditions $NEC_2$ and the $DEC_1 $, are satisfied.

We will not show all energy conditions inside the event horizon, neither for this case nor for the cases of the next solutions, because all components of the metric do not change their sign when analyzed inside the event horizon. So we will always have one temporal and three spatial coordinates. It follows that the energy conditions outside the horizon are the same as the energy conditions inside the horizon. So in all following cases, i.e. for solutions of $\mathbb{Q}=0$, we will show only the energy conditions outside the event horizon.


\subsection{Second solution of regular black holes with zero non-metricity scalar, $\mathbb{Q}=0$} \label{sec5D}

Now let us adopt $b(r)=-2\ln[1 - 2Mr/(r^2 + A^2)]$. In the case of regular black holes the non-metricity scalar \eqref{scalarQ} for this case is, 
\begin{equation}
0=-\frac{2 e^{-b(r)} \left[r a'(r)+1\right]}{r^2},
\end{equation}
whose solution is, 
\begin{equation}
a(r)=-\ln \left(\frac{r}{r_0}\right),
\end{equation}
where $r_0$ is a constant. We will use $r_0=1$. 

In this case, the Kretschmann scalar diverges when we take the $r\to0$ limit. 
At the horizon radius the Kretschamnn scalar depends only on the constants. And when we evaluate it at $r\to\infty$, this scalar has an asymptotically flat behavior. 

The equation of the geodesic is described by,
\begin{equation}
\dot{r}^2   = -\frac{1}{r}\left(1-\frac{2 M r}{A^2+r^2}\right)^2 \left(E^2 r^2+\frac{l^2}{r}\right)+\epsilon  \left(1-\frac{2 M r}{A^2+r^2}\right)^2,\label{geod_bnr2}
\end{equation}
where $l$ is the angular momentum and $E $ is the energy. 

By analyzing the geodesic equation \eqref{geod_bnr2} for $r$ very large, we obtain a solution for $r(s)$ that diverges at infinity at $s\to\infty$, suggesting that the geodesic equation is extensible at future infinity. When we analyze the equation of the geodesic for $r(s)\to 0$, we find that the equation of the geodesic is not extensible for this condition. This can also be checked using the Kretschmann scalar, since this scalar diverges in the limit of $r(s)\to 0$. We now analyze the geodesic equation for the horizon radius, which in this case is given by $r\to M+\sqrt{M^2-A^2}$. The solution we obtain for $r(s)$ shows us that this equation is extensible when evaluated in this limit. Therefore, this spacetime is geodesically complete.

Considering $f(\mathbb{Q})=0$, $f_{Q}(r)=1$ and $f_{QQ}(r)=0$. Now the moment-energy tensor has the components as,
\begin{eqnarray}
 \kappa ^2 \rho (r) &=& \frac{4 M \left(2 A^4+A^2 r (2 r-3 M)+M r^3\right)}{r \left(A^2+r^2\right)^3}, \\
 \kappa ^2p_r(r) &=& -\frac{1}{r^2}, \\
 \kappa ^2p_t(r) &=& \frac{\left(A^2+r (r-2 M)\right) \left(A^4+2 A^2 r (r-3 M)+r^3 (2 M+r)\right)}{4 r^2 \left(A^2+r^2\right)^3}.
\end{eqnarray}
Performing the limit $r\to 0$ we have an indeterminacy for the component $\rho(r)\to 0$, for the other components we have divergences $p_r\to-\infty$ and $p_t\to+\infty$. And for the $r\to-\infty$ limit the components $\rho(r)$, $p_r(r)$ and $p_t(r)$ are null.
\par
Then, the representation of the Ricci scalar is,
\begin{equation}
\overset{C}{R}=-\frac{A^6+3 A^4 r (8 M+r)+3 A^2 r^2 \left(-12 M^2+8 M r+r^2\right)+12 M^2 r^4+r^6}{2 r^2 \left(A^2+r^2\right)^3}.\label{RBNR1}    
\end{equation}
And the expression for the total derivative term \eqref{boundary},
\begin{equation}
    B_Q=\frac{A^6+3 A^4 r (8 M+r)+3 A^2 r^2 \left(-12 M^2+8 M r+r^2\right)+12 M^2 r^4+r^6}{2 r^2 \left(A^2+r^2\right)^3}.\label{B_BNR1}
\end{equation}
Since the non-metricity scalar is zero, the terms of $R$ \eqref{RBNR1} differ from $B_Q$ \eqref{B_BNR1} by a negative sign, which guarantees our results of \eqref{scalar_Ric2}. For this solution, since $\mathbb{Q}=0$ and by the direct relation that $\overset{C}{R}=-B_Q$, we find that for the point at $r=0$ a singularity is evident. So we can conclude that for the solutions of $\mathbb{Q}=0$ the solutions remain singular even if the metric function $b(r)$ is regular. In contrast to the linear theory, it is shown that spacetime is regular even if the non-metricity scalar is singular and the metric functions $a(r)$ and $b(r)$ are regular.
\par
Thus, the energy conditions outside the horizon are,
\begin{eqnarray}
&&NEC_{1}=SEC_{1}=WEC_{1}=	\frac{1}{r^2\kappa^{2}}\left[\frac{4 M r \left(2 A^4+A^2 r (2 r-3 M)+M r^3\right)}{\left(A^2+r^2\right)^3}-1\right]\ge0,\\
&&NEC_{2}=SEC_{2}=WEC_{2}=\frac{A^6+3 A^4 r (8 M+r)+3 A^2 r^2 \left(-12 M^2+8 M r+r^2\right)+12 M^2 r^4+r^6}{4 \kappa ^2 r^2 \left(A^2+r^2\right)^3}\ge0,\\
&&SEC_{3}=\frac{-A^6+A^4 r (8 M-3 r)+A^2 r^2 \left(-12 M^2+8 M r-3 r^2\right)+4 M^2 r^4-r^6}{2 \kappa ^2 r^2 \left(A^2+r^2\right)^3}\ge0,\\
&&DEC1=		\frac{1}{\kappa^{2} r^2}\left[ \frac{4 M r \left(2 A^4+A^2 r (2 r-3 M)+M r^3\right)}{\left(A^2+r^2\right)^3}+1 \right]\ge0,\\
&&DEC_{2}=\frac{-A^6+A^4 r (40 M-3 r)+A^2 r^2 \left(-60 M^2+40 M r-3 r^2\right)+20 M^2 r^4-r^6}{4 \kappa ^2 r^2 \left(A^2+r^2\right)^3}\ge0,\\
&&DEC_{3}=WEC_{3}=\frac{4 M \left(2 A^4+A^2 r (2 r-3 M)+M r^3\right)}{\kappa ^2 r \left(A^2+r^2\right)^3}\geq0.
\end{eqnarray}
\par
We note that the energy conditions $NEC_1$, $SEC_3$, and $DEC_2$ are violated for very small $r$, while the energy conditions $NEC_2$, $DEC_1$ and $DEC_3$ are satisfied. And for $r$ very large, the energy conditions $NEC_1$, $SEC_3$, $DEC_2$ are violated, while the energy conditions $NEC_2$, $DEC_1$, and $DEC_3$ are satisfied.


\section{Black-Bounce solutions in the $f(\mathbb{Q})$ theory }
\label{sec6}
We will now examine the black-bounce solutions in $f(\mathbb{Q})$ theory. Complementing this, we develop the solutions for STEGR starting from the Simpson-Visser interpretation and then for the cases where we consider the non-metricity scalar to be zero. We note that the equations of the components of the momentum-energy tensor, which we obtain in the case of STEGR, are the same as in the case of the black-bounce of general relativity. For each case we address, we calculate the energy conditions and discuss their authenticity.
\par
Black-bounce type solutions are obtained by considering the static spherically symmetric line element \eqref{m_bb}, recall
\begin{equation*}
ds^2=e^{a(r)}dt^2-e^{b(r)}dr^2-\Sigma(r)^2\left[d\theta^{2}+\sin^{2}\left(\theta\right)d\phi^{2}\right]\label{ltb}\,,
\end{equation*}
whose the evolutionary equations are given by (\ref{eq_field_BB_fQ}-\ref{eq_field_BB_fQ2}).
\subsection{Solutions of Black-Bounce in Symmetric teleparallel theory equivalent to general relativity (STTEGR)}\label{sec6A}

Here, too, we can obtain black-bounce solutions in STTEGR with the help of the evolutionary equations (\ref{eq_field_BB_fQ}-\ref{eq_field_BB_fQ2}) $f(\mathbb{Q})=\mathbb{Q}$, $f_{Q}(r)=1$ and $f_{QQ}(r)=0$.
\begin{eqnarray}
\frac{e^{-b(r)} \left[\Sigma (r) \left(b'(r) \Sigma '(r)-2 \Sigma ''(r)\right)+e^{b(r)}-\Sigma '(r)^2\right]}{\Sigma (r)^2}=&&\kappa ^2 \rho (r),\label{eqf_STTEGRBB}\\
-\frac{e^{-b(r)} \left[\Sigma (r) a'(r) \Sigma '(r)-e^{b(r)}+\Sigma '(r)^2\right)}{\Sigma (r)^2}=&&\kappa ^2 p_r(r),\\
\frac{e^{-b(r)} \left(-2 a'(r) \Sigma '(r)-\Sigma (r) \left(2 a''(r)-a'(r) b'(r)+a'(r)^2\right)+2 b'(r) \Sigma '(r)-4 \Sigma ''(r)\right]}{4 \Sigma (r)}=&&-\kappa ^2 p_t(r).\label{eqf_STTEGRBB1}
\end{eqnarray}
\par
Using the symmetry where $a(r)=-b(r)$, $\Sigma(r)=\sqrt{L^2_0+r^2}$ and $b(r) =-\ln{\big[1-(2M/\Sigma(r))\big]}$, we find the following components, 
\begin{eqnarray}
&&\rho(r)	=-\frac{L_0^2 \left(\sqrt{{L_0}^2+r^2}-4 M\right)}{\kappa ^2 \left(L_0^2+r^2\right)^{5/2}},\label{rho3}\\
&&p_{r}(r)	=-\frac{L_0^2}{\kappa ^2 \left(L_0^2+r^2\right)^2},\label{pr3}\\
&&p_{t}(r)	=\frac{ L_0^2 \left(\sqrt{L_0^2+r^2}- M\right)}{\kappa ^2 \left(L_0^2+r^2\right)^{5/2}}.\label{pt3}
\end{eqnarray}

Note that the expressions (\ref{rho3}-\ref{pt3}) are the same black-bounce solutions obtained in the Simpson-Visser model of general relativity \cite{visser2}. So, from symmetric teleparallel theory we get the same black-bounce solutions as with the general relativity theory. 
\par
Now we have that the non-metricity scalar is given by,
\begin{equation}
    \mathbb{Q}=-\frac{2 r^2}{\left({L_0}^2+r^2\right)^2}.\label{QBB}
\end{equation}
Thus, the Ricci scalar is,
\begin{equation}
   \overset{C}{R}=\frac{2 {L_0}^2 \left(\sqrt{{L_0}^2+r^2}-3 M\right)}{\left({L_0}^2+r^2\right)^{5/2}}.\label{RBB}
\end{equation}
\par
We get the following expression for the boundary term,
\begin{equation}
  B_Q= \frac{6 {L_0}^2 M}{\left({L_0}^2+r^2\right)^{5/2}}-\frac{2}{{L_0}^2+r^2},\label{B_BB}
\end{equation}
and from the expressions \eqref{QBB}, \eqref{RBB} and \eqref{B_BB} it follows again that the relation \eqref{scalar_Ric2} is satisfied.
We can verify that  \eqref{QBB}, \eqref{RBB} and \eqref{B_BB}, are always finite for any value of the radial coordinate $r$.

We note here that, unlike solutions for black holes and regular black holes, both the non-metricity scalar $\mathbb{Q}$ and the boundary term $B_Q$ are always regular throughout spacetime. This was not the case even for the Minkowski metric.

\par
\subsection{First black-bounce solution with zero non-metricity scalar, $\mathbb{Q}=0$} \label{sec6B}

Black-bounce solutions can be obtained by means of the following imposition $\Sigma(r)=\sqrt{L^2_0+r^2}$, where we use $b(r) =-2\ln{[1-(2M/\Sigma(r))]}$, this way we get the arbitrary function $a(r)$. So we can make $\mathbb{Q}=0$ to get the function $a(r)$. The equation \eqref{scalarQ} becomes, 
\begin{equation}
    -\frac{2e^{-b(r)}\Sigma'(r)\left[\Sigma(r)a'(r)+\Sigma'(r)\right]}{\Sigma^2(r)}=0.
\end{equation}
so we have the following solution for $a(r)$, 
\begin{equation}
a({r})=-\frac{1}{2}\ln\left(\frac{L_{0}^{2}+r^{2}}{r_0^2}\right).\label{a1}
\end{equation}
where $r_0$ is a constant. We will use $r_0=1$. 

We verified from the solution \eqref{a1} and the parameter $b(r)$, that the Kretschamnn scalar behaves regularly in spacetime. For we note that the Kretschmann scalar is regular for any value of $r$ and that in the middle, i.e. in the limit $r(r)\to0$, this scalar behaves like a constant which depends only on $m$ and $L_0$. At the radius of the horizon the scalar also behaves like a constant. And at $r\to\infty$ this scalar is asymptotically flat.

The equation of the geodesic \eqref{geodesic} for this case is,
\begin{equation}
   \left(\frac{dr}{d s}\right)^2= -\frac{1}{\sqrt{L_0^2+r^2}}\left(1-\frac{2 M}{\sqrt{L_0^2+r^2}}\right)^2 \left(E^2 \left(L_0^2+r^2\right)+\frac{l^2}{\sqrt{L_0^2+r^2}}\right)+\epsilon  \left(1-\frac{2 M}{\sqrt{L_0^2+r^2}}\right)^2, \label{geod_bb} 
\end{equation}
where $l$ is the angular momentum and $E$ is the energy.

We note that the geodesic equation \eqref{geod_bb}, when analyzed for $r\to\infty$, yields a solution for $r(s)$ that diverges at $s\to\infty$, implying that the geodesic equation is extensible to future infinity. This equation is also extensible when we evaluate for very small $r(s)$. We also note that the solution we obtain in this case gives us a geodesic equation that is also extensible to $r(s)\to 0$. This result can be confirmed by the Kretschmann scalar, which we discussed earlier. Now analyzing the geodesic equation for the horizon radius $r_H=\sqrt{4 M^2-L_0^2}$, we find from the solution we obtain for $r(s)$ that the equation \eqref{geod_bb} is also extensible at $r_H$.  Therefore, this space-time is geodesically complete. 

To simplify the equations for the components of the momentum-energy tensor, we assume the following functions: $f(\mathbb{Q})=0$, $f_{Q}(r)=1$, and $f_{QQ}(r)=0$. Thus, the components of the moment-energy tensor resulting from the expressions (\ref{eq_field_BB_fQ}-\ref{eq_field_BB_fQ2}) and \eqref{a1} are
\begin{eqnarray}
&&\rho(r)	=-\frac{L_0^4+L_0^2 \left[8 M \left(M-\sqrt{L_0^2+r^2}\right)+r^2\right]-4 M^2 r^2}{\kappa ^2 \left(L_0^2+r^2\right)^3},\label{rho32}\\
&&p_{r}(r)	=-\frac{1}{\kappa ^2 \left(L_0^2+r^2\right)},\label{ptr32}\\
&&p_{t}(r)	= \frac{2 L_0^4+L_0^2 \left(8 M \left(M-\sqrt{L_0^2+r^2}\right)+3 r^2\right)-4 M^2 r^2+r^4}{4 \kappa ^2 \left(L_0^2+r^2\right)^3} .\label{pt32}
\end{eqnarray}
\par
Let us now analyze the limit of the components $\rho(r)$, $p_r(r)$, and $p_t(r)$ to ensure that the solutions are free of divergences in spacetime. In this way, we can derive the necessary fluid for the case when $\mathbb{Q}=0$. Then, taking the limit of $r\rightarrow\infty$ we will have ($\rho(r),p_r(r)$,$p_t$(r))$\rightarrow 0$, and for $r\rightarrow0$,
\begin{eqnarray}
&&\rho(r)\rightarrow\frac{8 M \left(\sqrt{L_0^2}-M\right)-L_0^2}{\kappa ^2 L_0^4},\\
&&p_{r}(r)\rightarrow-\frac{1}{\kappa ^2 L_0^2},\\
&&p_{t}(r)\rightarrow \frac{-4 \sqrt{L_0^2} M+L_0^2+4 M^2}{2 \kappa ^2 L_0^4}.
\end{eqnarray}
Note that we cannot determine the liquid in the infinite future.
At the origin of the system, we find that the density components and the radial and tangential pressures depend on a certain value of $L_0$.
\par
Now the ricci scalar has the shape, 
\begin{equation}
  \overset{C}{R}= \frac{2 L_0^4+L_0^2 \left(-24 M \sqrt{L_0^2+r^2}+24 M^2+r^2\right)-r^2 \left(12 M^2+r^2\right)}{2 \left(L_0^2+r^2\right)^3}.\label{RBB1}
\end{equation}
With the boundary term \eqref{boundary} being,
\begin{equation}
 B_Q=\frac{-2 L_0^4+L_0^2 \left(24 M \left(\sqrt{L_0^2+r^2}-M\right)-r^2\right)+12 M^2 r^2+r^4}{2 \left(L_0^2+r^2\right)^3},\label{B_BB1}
\end{equation}
so since $\mathbb{Q}=0$, the scalar \eqref{RBB1} and the boundary term \eqref{B_BB1} are relations satisfying the equation \eqref{scalar_Ric2}. Now spacetime is always regular. This time the scalars give us the correct interpretation that spacetime remains regular for any value of the radial coordinate.
\par
By means of the equations (\ref{rho32}-\ref{pt32}), we can develop the energy conditions for regions outside the horizon being given by,
\begin{eqnarray}
&&NEC_{1}=SEC_{1}=WEC_{1}=\frac{-2 L_0^4+L_0^2 \left(8 M \sqrt{L_0^2+r^2}-8 M^2-3 r^2\right)+4 M^2 r^2-r^4}{\kappa ^2 \left(L_0^2+r^2\right)^3}\ge0\\
&&NEC_{2}=SEC_{2}=WEC_{2}=\frac{-2 L_0^4+L_0^2 \left(24 M \sqrt{L_0^2+r^2}-24 M^2-r^2\right)+12 M^2 r^2+r^4}{4 \kappa ^2 \left(L_0^2+r^2\right)^3}\ge0,\\
&&SEC_{3}=	\frac{-2 L_0^4+L_0^2 \left(8 M \sqrt{L_0^2+r^2}-8 M^2-3 r^2\right)+4 M^2 r^2-r^4}{2 \kappa ^2 \left(L_0^2+r^2\right)^3} \ge0,\\
&&DEC1=	\frac{L_0^2 \left(8 M \sqrt{L_0^2+r^2}-8 M^2+r^2\right)+4 M^2 r^2+r^4}{\kappa ^2 \left(L_0^2+r^2\right)^3} \ge0,\\
&&DEC_{2}=	\frac{-6 L_0^4+L_0^2 \left(40 M \sqrt{L_0^2+r^2}-40 M^2-7 r^2\right)+20 M^2 r^2-r^4}{4 \kappa ^2 \left(L_0^2+r^2\right)^3} \ge0,\\
&&DEC_{3}=WEC_{3}=-\frac{L_0^4+L_0^2 \left(8 M \left(M-\sqrt{L_0^2+r^2}\right)+r^2\right)-4 M^2 r^2}{\kappa ^2 \left(L_0^2+r^2\right)^3} \geq0.
\end{eqnarray}
\par
In this case only $NEC_2$ is violated, the other energy conditions are satisfied for very small $r$. The energy conditions $DEC_1$ and $DEC_3$ are satisfied for very large $r$, while the other energy conditions are violated.


\subsection{Second black-bounce solution with zero non-metricity scalar, $\mathbb{Q}=0$} \label{sec6C}

For this case we will consider $b(r) =-2\ln{[1-2Mr/(L_0^2+r^2)]}$ and $\Sigma(r)=\sqrt{L^2_0+r^2}$, the solution to $a(r)$ is given by, 
\begin{equation}
a({r})=-\frac{1}{2}\ln\left(\frac{L_{0}^{2}+r^{2}}{r_0^2}\right),\label{a1_BB2}
\end{equation}
where $r_0$ is a constant. We will use $r_0=1$. 

We note that the Kretschamnn scalar for this case also shows a regular behavior in spacetime for any value of $r$. In the limit of $r\to0$ this scalar behaves like a constant and is proportional to $K\sim L_0^{-4}$. In the horizon radius this scalar also behaves like a constant which depends on $m$ and $L_0$. And in the limit of $r\to\infty$ the solution shows an asymptotically flat behavior.

The equation of the geodesic \eqref{geodesic} is now described by,
\begin{equation}
   \left(\frac{dr}{d s}\right)^2 =-\frac{1}{\sqrt{L_0^2+r^2}}\left(1-\frac{2 M r}{L_0^2+r^2}\right)^2 \left(E^2 \left(L_0^2+r^2\right)+\frac{l^2}{\sqrt{L_0^2+r^2}}\right)+\epsilon  \left(1-\frac{2 M r}{L_0^2+r^2}\right)^2,\label{geod_bb2} 
\end{equation}
where $l$ is the angular momentum and $E$ is the energy.

We note that the solution obtained from the equation of the geodesics \eqref{geod_bb2}, if analyzed very large for $r$, is extensible to future infinity. This equation can also extensible to very small $r$. We also note that the geodesic equation in this case is extensible to $r(s)\to 0$. This can be confirmed with our result for the Kretschmann scalar described earlier. For the radius of the horizon $r\to r_H$, which in this case is given by $r_H=M+\sqrt{M^2-L^2}$, we note that the geodesic equation is also extensible. Therefore this space-time is geodesically complete.

Assuming that the following functions are $f(\mathbb{Q})=0$, $f_{Q}(r)=1$, and $f_{QQ}(r)=0$, we obtain the following components of the momentum-energy tensor from the expressions (\ref{eq_field_BB_fQ}-\ref{eq_field_BB_fQ2}) and \eqref{a1_BB2}, which are now described as follows
\begin{eqnarray}
&&\rho(r)	= --\frac{L_0^6+2 L_0^4 r (r-6 M)+L_0^2 r^2 \left(16 M^2-12 M r+r^2\right)-4 M^2 r^4}{\kappa ^2 \left(L_0^2+r^2\right)^4},\label{rho33}\\
&&p_{r}(r)	=-\frac{1}{\kappa ^2 \left(L_0^2+r^2\right)},\label{pr33}\\
&&p_{t}(r)	= \frac{\Big[L_0^2+r (r-2 M)\Big] \Big[2 L_0^4+L_0^2 r (3 r-8 M)+r^3 (2 M+r)\Big]}{4 \kappa ^2 \left(L_0^2+r^2\right)^4}.\label{pt33}
\end{eqnarray}
\par
The asymptotic limits of $r\rightarrow\infty$ for the components are null ($\rho(r),p_r(r)$,$p_t$(r))$\rightarrow 0$, and for $r\rightarrow0$ they reduce to,
\begin{eqnarray}
&&\rho(r)\rightarrow-\frac{1}{\kappa ^2 L_0^2},\\
&&p_{r}(r)\rightarrow-\frac{1}{\kappa ^2 L_0^2},\\
&&p_{t}(r)\rightarrow \frac{1}{2\kappa ^2 L_0^2}.
\end{eqnarray}
At the origin of the system, we see that the density components and the radial and tangential pressures depend on a certain value of $L_0$.

For this case we have the following Ricci scalar,
\begin{equation}
  \overset{C}{R}= \frac{2 L_0^6+3 L_0^4 r (r-12 M)+12 L_0^2 M r^2 (4 M-3 r)-r^4 \left(12 M^2+r^2\right)}{2 \left(L_0^2+r^2\right)^4}.\label{RBB2}
\end{equation}
\par
Then we have,
\begin{equation}
 B_Q=\frac{-2 L_0^6+3 L_0^4 r (12 M-r)+12 L_0^2 M r^2 (3 r-4 M)+12 M^2 r^4+r^6}{2 \left(L_0^2+r^2\right)^4},\label{B_BB2}
\end{equation}
again we have that the expressions \eqref{RBB2} and \eqref{B_BB2} satisfy the expression \eqref{scalar_Ric2}.  The space-time is always regular.
\par
From the components of (\ref{rho33}-\ref{pt33}), the energy conditions for regions outside the horizon are, 
\begin{eqnarray}
&&NEC_{1}=SEC_{1}=WEC_{1}= \frac{-2 L_0^6+L_0^4 r (12 M-5 r)-4 L_0^2 r^2 \left(4 M^2-3 M r+r^2\right)+4 M^2 r^4-r^6}{\kappa ^2 \left(L_0^2+r^2\right)^4} \ge0,\\
&&NEC_{2}=SEC_{2}=WEC_{2}= \frac{-2 L_0^6+3 L_0^4 r (12 M-r)+12 L_0^2 M r^2 (3 r-4 M)+12 M^2 r^4+r^6}{4 \kappa ^2 \left(L_0^2+r^2\right)^4} \ge0,\\
&&SEC_{3}= -\frac{2 L_0^6+L_0^4 r (5 r-12 M)+4 L_0^2 r^2 \left(4 M^2-3 M r+r^2\right)-4 M^2 r^4+r^6}{2 \kappa ^2 \left(L_0^2+r^2\right)^4} \ge0,\\
&&DEC1=	 \frac{r \left(L_0^4 (12 M+r)+2 L_0^2 r \left(-8 M^2+6 M r+r^2\right)+4 M^2 r^3+r^5\right)}{\kappa ^2 \left(L_0^2+r^2\right)^4} \ge0,\\
&&DEC_{2}= -\frac{6 L_0^6+L_0^4 r (13 r-60 M)+4 L_0^2 r^2 \left(20 M^2-15 M r+2 r^2\right)-20 M^2 r^4+r^6}{4 \kappa ^2 \left(L_0^2+r^2\right)^4} \ge0,\\
&&DEC_{3}=WEC_{3}= -\frac{L_0^6+2 L_0^4 r (r-6 M)+L_0^2 r^2 \left(16 M^2-12 M r+r^2\right)-4 M^2 r^4}{\kappa ^2 \left(L_0^2+r^2\right)^4} \geq0.
\end{eqnarray}

We note that all energy conditions are violated for very small $r$. The energy conditions $NEC_1$, $SEC_3$ and $DEC_2$ are violated for too large $r$, while the other energy conditions are satisfied. 


\subsection{Third black-bounce solution with zero non-metricity scalar, $\mathbb{Q}=0$}\label{sec6D}

In the latter case we will assume $b(r) =-2\ln{[1-2M/\Sigma(r)]}$, but  with,
\begin{equation}
    \Sigma(r)=\sqrt{\left(L_0^2+r^2\right) \exp \left(\frac{r_0}{r_1^2+r^2}\right)},
\end{equation}
we use this proposal with the intention of obtaining black-bounce solutions satisfying $NEC_1$. The solution is,
\begin{equation}
a({r})=-\frac{r_0}{2 \left(r_1^2+r^2\right)}-\frac{1}{2}\ln\left(\frac{L_0^{2}+r^{2}}{r_3^2}\right),\label{a3}
\end{equation}
where $r_3$ is a constant. We will use $r_3=1$. 

The Kretschamnn scalar behaves in this case like a constant in the limit $r\to0$. In the horizon radius this scalar also behaves like a constant. However, when we evaluate the scalar for $r\to0$, we do not find asymptotically flat behavior.

\begin{eqnarray}
   && \left(\frac{dr}{d s} \right)^2= - \frac{e^{-\frac{r_0}{2 \left(r^2+r_1^2\right)}} }{\sqrt{L_0^2+r^2}}\left(1-\frac{2 M}{\sqrt{\left(L_0^2+r^2\right) e^{\frac{r_0}{r^2+r_1^2}}}}\right)^2 \left(\frac{l^2 e^{-\frac{r_0}{2 \left(r^2+r_1^2\right)}}}{\sqrt{L_0^2+r^2}}+E^2 \left(L_0^2+r^2\right) e^{\frac{r_0}{r^2+r_1^2}}\right)+\nonumber\\
   &&\phantom{\left(\frac{dr}{d\lambda} \right)^2=} +\epsilon  \left(1-\frac{2 M}{\sqrt{\left(L_0^2+r^2\right) e^{\frac{r_0}{r^2+r_1^2}}}}\right)^2\label{geod_bb3}
\end{eqnarray}
where $l$ is the angular momentum and $E$ is the energy.

The solution of the geodesic equation for $r(s)$   very large \eqref{geod_bb3}, is an extensible solution at future infinity. And the geodesic equation for $r(s)$ very small, is a solution that is also extensible. Now the solution of $r(s)$ given by \eqref{geod_bb3} is also extensible for $r(s)\to 0$.
We also note that this equation, analyzed at the radius of the horizon, is also extensible. Therefore, this spacetime is geodesically complete.

Considering again the functions $f(\mathbb{Q})=0$, $f_{Q}(r)=1$ and $f_{QQ}(r)=0$, the components of the moment-energy tensor are given by, 
\begin{eqnarray}
&&\rho(r)= \frac{1}{\kappa ^2 \left(L_0^2+r^2\right)^2}\left\{\left(L_0^2+r^2\right) e^{-\frac{r_0}{r^2+r_1^2}}-\frac{r^2 \left(-L_0^2 r_0+r^4-r^2 \left(r_0-2 r_1^2\right)+r_1^4\right)^2 \left(1-\frac{2 M}{\sqrt{\left(L_0^2+r^2\right) e^{\frac{r_0}{r^2+r_1^2}}}}\right)^2}{\left(r^2+r_1^2\right)^4}\right.\\
&& -\frac{2 \left(1-\frac{2 M}{\sqrt{\left(L_0^2+r^2\right) e^{\frac{r_0}{r^2+r_1^2}}}}\right)^2}{\left(r^2+r_1^2\right)^4 \left(\sqrt{\left(L_0^2+r^2\right) e^{\frac{r_0}{r^2+r_1^2}}}-2 M\right)} \times \Bigg[r^4 r_0 \left(r^4+r^2 \left(r_0-2 r_1^2\right)-3 r_1^4\right) \sqrt{\left(L_0^2+r^2\right) e^{\frac{r_0}{r^2+r_1^2}}}+ \nonumber \\
&& +2 M r^2 \left(r^2+r_1^2\right) \left(r^6+3 r^4 \left(r_1^2-r_0\right)+r^2 r_1^2 \left(r_0+3 r_1^2\right)+r_1^6\right)+\nonumber \\
&& +L_0^4 r_0 \left(\left(3 r^4+r^2 \left(r_0+2 r_1^2\right)-r_1^4\right) \sqrt{\left(L_0^2+r^2\right) e^{\frac{r_0}{r^2+r_1^2}}}+2 M \left(-3 r^4-2 r^2 r_1^2+r_1^4\right)\right)+ \nonumber\\
&& L_0^2\Bigg(\left(r^8+4 r^6 \left(r_0+r_1^2\right)+2 r^4 \left(r_0^2+3 r_1^4\right)+4 r^2 r_1^4 \left(r_1^2-r_0\right)+r_1^8\right) \sqrt{\left(L_0^2+r^2\right) e^{\frac{r_0}{r^2+r_1^2}}}+\nonumber\\
&& -2 M \left(r^2+r_1^2\right) \left(r^6+3 r^4 \left(2 r_0+r_1^2\right)+r^2 \left(3 r_1^4-2 r_0 r_1^2\right)+r_1^6\right)\Bigg)\Bigg]\Bigg\} \label{rho34},	\\
&&p_{r}(r)	=-\frac{e^{-\frac{r_0}{r_1^2+r^2}}}{\kappa ^2 \left({L_0}^2+r^2\right)},\label{pr34}\\
&&p_{t}(r)	= \frac{-4r^{8}\left(Mr_0-r_1^{2}\sqrt{\left(L_0^{2}+r^{2}\right)e^{\frac{r_0}{r^{2}+r_1^{2}}}}\right)+r^{6}\left(\left(3r_0^{2}-8r_0r_1^{2}+6r_1^{4}\right)\sqrt{\left(L_0^{2}+r^{2}\right)e^{\frac{r_0}{r^{2}+r_1^{2}}}}-4Mr_0\left(r_0-2r_1^{2}\right)\right)}{4\kappa^{2}\left(L_0^{2}+r^{2}\right)^{2}\left(r^{2}+r_1^{2}\right)^{4}\sqrt{\left(L_0^{2}+r^{2}\right)e^{\frac{r_0}{r^{2}+r_1^{2}}}}}\nonumber\\	
&& \frac{+4r^{4}r_1^{4}\left(\left(r_1^{2}-2r_0\right)\sqrt{\left(L_0^{2}+r^{2}\right)e^{\frac{r_0}{r^{2}+r_1^{2}}}}+3Mr_0\right)+r^{2}r_1^{8}\sqrt{\left(L_0^{2}+r^{2}\right)e^{\frac{r_0}{r^{2}+r_1^{2}}}}+r^{10}\sqrt{\left(L_0^{2}+r^{2}\right)e^{\frac{r_0}{r^{2}+r_1^{2}}}}+}{4\kappa^{2}\left(L_0^{2}+r^{2}\right)^{2}\left(r^{2}+r_1^{2}\right)^{4}\sqrt{\left(L_0^{2}+r^{2}\right)e^{\frac{r_0}{r^{2}+r_1^{2}}}}} \nonumber	\\
&&  \frac{+L_0^{4}r_0\left(\left(6r^{4}+r^{2}\left(3r_0+4r_1^{2}\right)-2r_1^{4}\right)\sqrt{\left(L_0^{2}+r^{2}\right)e^{\frac{r_0}{r^{2}+r_1^{2}}}}-4M\left(3r^{4}+r^{2}\left(r_0+2r_1^{2}\right)-r_1^{4}\right)\right)}{4\kappa^{2}\left(L_0^{2}+r^{2}\right)^{2}\left(r^{2}+r_1^{2}\right)^{4}\sqrt{\left(L_0^{2}+r^{2}\right)e^{\frac{r_0}{r^{2}+r_1^{2}}}}}+\nonumber\\
&& \frac{+2L_0^{2}\Bigg[\left(r^{8}+r^{6}\left(3r_0+4r_1^{2}\right)+r^{4}\left(3r_0^{2}-2r_0r_1^{2}+6r_1^{4}\right)+r^{2}\left(4r_1^{6}-5r_0r_1^{4}\right)+r_1^{8}\right)\sqrt{\left(L_0^{2}+r^{2}\right)e^{\frac{r_0}{r^{2}+r_1^{2}}}}}{4\kappa^{2}\left(L_0^{2}+r^{2}\right)^{2}\left(r^{2}+r_1^{2}\right)^{4}\sqrt{\left(L_0^{2}+r^{2}\right)e^{\frac{r_0}{r^{2}+r_1^{2}}}}}+\nonumber\\
&& -2M\left(r^{8}+4r^{6}\left(r_0+r_1^{2}\right)+2r^{4}\left(r_0^{2}+3r_1^{4}\right)+4r^{2}r_1^{4}\left(r_1^{2}-r_0\right)+r_1^{8}\right)\Bigg]\label{pt34} .	
\end{eqnarray}

We also check in this case that the condition \eqref{scalar_Ric2} is satisfied.

\begin{figure}[h]
\centering
\includegraphics[scale=0.75]{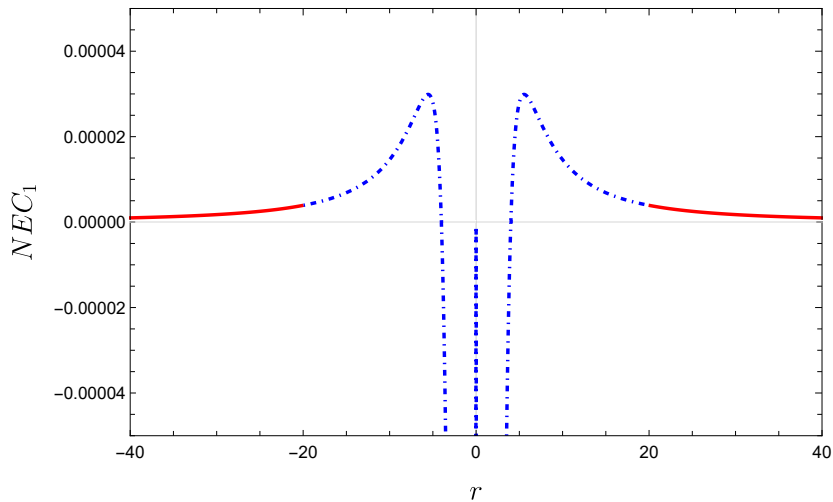}
\caption{\scriptsize{Graphical representation of the null energy condition \eqref{NEC} of spacetime, for the components $\rho$, $p_r$ and $p_t$, equations \eqref{rho34}, \eqref{pr34} and \eqref{pt34}, respectively. Where we consider $M=10$, $r_0=1$, $r_1=1$, and $L_0=1$ .} }
\label{fig1}
\end{figure}

The energy conditions resulting from (\ref{rho34}-\ref{pt34}) are reproduced in the figure \eqref{fig1} from the null energy condition (because they turn out to be quite extensive and challenging ), where the regions outside the event horizon are represented by the solid red curve and the regions inside the horizon by the dotted-trace curves with blue color.

\newpage
From the energy conditions (\ref{rho34}-\ref{pt34}) we see that $NEC_1$, as shown in \eqref{fig1}, is satisfied for
$r_1\rightarrow \infty$ and from $-r_1$ to $- \infty$. Here $r_1$ is a value inside the horizon violated near the throat. However, in general relativity, most models described by the relation $\Sigma''(r)/\Sigma(r) > 0$ imply that $NEC_1$ or $NEC_2$ are always violated. Unlike our model, where this condition is satisfied~\cite{ManuelSILVA}. When we consider $r$ very large, we find that $NEC_2$ is also satisfied, as is $DEC_1$. On the other hand, $SEC_3$ and $DEC_2$ are violated. We note, however, that the energy condition $DEC_3$ may or may not be violated by the choice of constants.


\section{Conclusion}\label{sec7}
\par
In this paper we study solutions of black holes, regular black holes and black-bounce spacetime in the theory $f(\mathbb{Q})$ using the coincident gauge. This theory is an extension of the symmetric teleparallel theory, where we use arbitrary functions of the non-metricity scalar $\mathbb{Q}$ in the action. We analyze the linear cases of our solutions and the solutions with the constraint $\mathbb{Q}=0$. We also calculate the Kretschmann scalar to observe regularities of spacetime, and briefly discuss the causal structure of spacetime using geodesics and the energy conditions of regular black holes and black-bounce.
We also calculate the Kretschmann scalar to observe regularities of spacetime and briefly discuss the causal structure of spacetime using geodesics, and the energy conditions of the solutions of regular black holes and black-bounce.

In the black hole solutions discussed in \eqref{sec4}, we obtain the linear case where we analyze the Kretschmann scalar and find a divergence at $r = 0$, this is a singularity. For the causal structure, we note that the geodesic equation is not extensible along spacetime in this case. This agrees with the result we obtained, for example, with the Kretschmann scalar for $r\rightarrow0$. For the causal structure, we note that the geodesic equation is not extensible along spacetime in this case. This agrees with the result we obtained with the Kretschmann scalar for $r\rightarrow0$. We solve with $\mathbb{Q}=0$~\eqref{subsec32}, use the metric function given by $b(r) =-2\ln{[1-2M/\Sigma(r)]}$, and obtain $a(r) =-\ln \left(r/r_0\right)$. So we find that spacetime is not geodesically complete. And by the Kretschmann scalar we have a spacetime which is not regular.

We also obtain the solutions for regular black holes \eqref{sec5}. For the Bardeen model, i.e., the linear case, we analyze the Kretschmann scalar and verify that our solutions are regular throughout spacetime. In the next case \eqref{subsec5A}, we use the metric function $b(r)=-2\ln\big[1 - 2Mr^2/(r^2 + q^2)^{3/2}\big]$ in the constraint where $\mathbb{Q}=0$, and consequently find the free function $a(r)=-\ln \left(r/r_0\right)$. Using the metric functions $b(r)$ and $a(r)$, we find that the geodesics are extensible over the entire spacetime. Therefore, this spacetime is geodesically complete. We calculate the energy density and the radial and tangential pressures. We find that this fluid, or rather the radial and tangential pressure components, diverge at the origin of spacetime. In the origin the density is determined by a certain value of the parameter $q$ and in the future infinity all components are equal to zero. This is confirmed by the Kretschmann scalar, which diverges at $r\to 0$ and shows an asymptotically  asymptotic flat behavior for $r\to \infty$. The energy conditions show that only $NEC_2$ is satisfied for very small $r$, while for very large $r$ only $NEC_2$ and $DEC_1$ are satisfied. In the second case of RBH \eqref{subsec5B}, with the non-metricity scalar zero, we use the metric function given by $b(r)=-2\ln[1 - 2Mr/(r^2 + A^2)]$, so that we find $a(r)=-\ln \left(r/r_0\right)$. Therefore, we can prove that the spacetime of this solution is geodesically complete.
On the other hand, even if the non-metricity scalar is singular, we have that the metric functions $a(r)$ and $b(r)$ are regular, and consequently it shows that spacetime is regular. The energy conditions $NEC_1$, $SEC_3$ and $DEC_2$ are violated for very small $r$, while the other energy conditions are satisfied. And the energy conditions for very large $r$ show for this case that $NEC_1$, $SEC_3$ and $DEC_2$ are violated, while the others are satisfied.

In section~\eqref{sec6} we study the solutions in the context of black-bounce. We verify for the case (STEGR) in~\eqref{sec6A} that the~(\ref{rho3}-\ref{pt3}) are the same solutions as in the Simpson-Visser model of GR.
And we also note that the non-metricity scalar \eqref{QBB} and the boundary term \eqref{B_BB} are regular throughout spacetime. This is in contrast to solutions for black holes and regular black holes, where we do not have a regular non-metricity scalar throughout spacetime. In the first black-bounce case with $\mathbb{Q}=0$ \eqref{sec6B}, we used the metric function $b(r) =-2\ln{[1-(2M/\Sigma(r))]}$ and $\Sigma(r)=\sqrt{L^2_0+r^2}$ to obtain the free function $a(r)=-\frac{1}{2}\ln\left(\frac{L_{0}^{2}+r^{2}}{r_0^2}\right)$, which is regular over spacetime. From the functions $a(r)$ and $b(r)$ in this case, we verify that the Kretschamnn scalar has regular behavior along spacetime. And we also examine the causal structure and find that this space-time is geodesically complete. We calculate the components of the fluid associated with this case, and by analyzing these components at the origin, we find that this fluid depends on constants such as $L_0$. And with infinite future all components are zero. Checking the energy conditions for this fluid, we see that only $NEC_2$ is satisfied, while the others are violated for $r$ very small.  And for very large $r$ only $DEC_1$ and $DEC_3$ are satisfied. 

In the second black-bounce solution with $\mathbb{Q}=0$~\eqref{sec6C}, we use the metric function, given by $b(r) =-2\ln{\bigg[1-\frac{2Mr}{(L_0^2+r^2)}\bigg]}$, where we find $a(r)=-\frac{1}{2}\ln\left(\frac{L_{0}^{2}+r^{2}}{r_0^2}\right)$. Using the metric functions in this case, we find that the Kretschamnn scalar has a regular behavior along spacetime for any value of $r$. Also in this case it is a complete geodesic spacetime. From the analysis of the energy conditions for density and radial and tangential pressure, we see that all energy conditions are violated for very small values of $r$. While only $NEC_1$, $SEC_3$ and $DEC_2$ are satisfied for very large $r$.

In~\eqref{sec6D} we have developed the third black-bounce solution for $\mathbb{Q}=0$. But now we use a special case with $\Sigma(r)=\sqrt{\left(L_0^2+r^2\right) \exp \left(\frac{r_0}{r_1^2+r^2}\right)}$ with the metric function $ b(r) =-2\ln{[1-2M/\Sigma(r)]}$, and we now have the following function $a(r)=-\frac{r_0}{2 \left (r_1^2+r^2\right)}-\frac{1}{2}\ln\left (\frac{L_0^{2}+r^{2}}{r_3^2}\right)$. This case shows that the Kretschamnn scalar is regular throughout spacetime. We also highlight the geodesics where we find that spacetime is geodesically complete. The energy conditions for this model are extremely challenging, so we constructed a graph of $NEC_1$ to observe its behavior. On this occasion we conclude that this energy condition is satisfied. Unlike in the black-bounce case of GR, where this energy condition is violated. 

We are aware of the challenges of the theory $f(\mathbb{Q})$ and that it is a theory that has only recently been formulated. On the other hand, although we find many works on this formulation, we want to use this theory to study other topics. In the future, we plan to use $f(\mathbb{Q})$ theory to study other approaches such as black hole thermodynamics, quasi-normal modes, black hole shadows, gravitational waves, applications in Kerr metrics, and perturbation theory. Moreover, we also see a possibility of using new gauges in this theory to apply in the context of black-bounce, for example.

\vspace{1cm}

{\bf Acknowledgement}:  M. E. R. thanks CNPq for partial financial support.  This study was supported in part by the Coordenção de Aperfeioamento de Pessoal de Nível Superior - Brazil (CAPES) - Financial Code 001. We would also like to thank Professor Gonzalo J. Olmo for discussions that were helpful for this article.  


%

\end{document}